\documentclass[journal]{IEEEtran}
\usepackage{amsfonts}
\usepackage{amsmath}
\usepackage{amssymb}
\usepackage{cite}
\usepackage{dsfont}
\usepackage{latexsym}
\usepackage{subfigure}
\usepackage{times}
\usepackage{url}
\usepackage{verbatim}

\usepackage{graphicx}
\newtheorem{theorem}{Theorem}

\newtheorem{algorithm}[theorem]{Algorithm}

\newcommand{\figref}[1]{{Fig.}~\ref{#1}}

\def\bb0{{\mathbb{0}}}

\def\bb{{\mathbf{b}}}

\def\bh{{\mathbf{h}}}

\def\bq{{\mathbf{q}}}

\def\bt{{\mathbf{t}}}

\def\bw{{\mathbf{w}}}
\def\bx{{\mathbf{x}}}

\def\bz{{\mathbf{z}}}
\def\b0{{\mathbf{0}}}

\def\bA{{\mathbf{A}}}
\def\bB{{\mathbf{B}}}

\def\bD{{\mathbf{D}}}

\def\bQ{{\mathbf{Q}}}

\def\bW{{\mathbf{W}}}
\def\bX{{\mathbf{X}}}
\def\bY{{\mathbf{Y}}}

\def\sf0{{\mathsf{0}}}

\def\rmM{\mathrm{M}}

\def\rmT{\mathrm{T}}

\def\rmc{{\mathrm{c}}}
\def\rmd{{\mathrm{d}}}

\def\rmj{{\mathrm{j}}}

\def\rm0{{\mathrm{0}}}

\def\kron{\otimes}

\usepackage[]{hyperref}
\hypersetup{pdftex,colorlinks=true,allcolors=blue}
\usepackage{hypcap}
\usepackage{enumitem}
\usepackage[usenames, dvipsnames]{color}

\newcommand{\comm}{\mathrm{c}}

\newcommand{\Es} {{\mathcal{E}_{\mathrm{s}}} }

\newcommand{\e}[1]{{\mathbb E}\left[ #1 \right]}

\newcommand{\mrad}{\mathrm{r}}

\newcommand{\mcom}{\comm}
\newcommand{\mr}{\mrad}

\newcommand{\mI}{\mathrm{si}}

\newcommand{\mrx}{\mathrm{RX}}

\newcommand{\Ts}{T_{\mathrm{s}}}

\newcommand{\txm}{\mathrm{TX}}

\newcommand{\thm}{\mathrm{th}}

\usepackage{graphicx}
 \usepackage{multirow}
 \usepackage[outdir=Fig/]{epstopdf}
\usepackage{amssymb}
\usepackage{algorithm}
\usepackage{algorithmic}
\usepackage{psfrag}

\usepackage{lipsum}
\usepackage{fancyhdr}
\usepackage{datetime}

\title{JCR70: A Low-Complexity Millimeter-Wave Proof-of-Concept Platform for A Fully-Digital MIMO Joint Communication-Radar}

\author{{Preeti~Kumari, {\it Student Member, IEEE} Amine~Mezghani, {\it Member, IEEE}, and~Robert~W.~Heath,~Jr., {\it Fellow, IEEE}}
\thanks{ Preeti Kumari and Robert W. Heath Jr. are with the Wireless Networking and Communications Group, The University of Texas at Austin, TX 78712-1687, USA (e-mail: \{preeti\_kumari, rheath\}@utexas.edu). Amine Mezghani is with University of Manitoba, MB R3T 5V6, Canada (email: amine.mezghani@umanitoba.ca).

We would like to thank National Instruments' Wireless Lead User Team for their assistance with the mmWave communication testbed.
This material is based upon work supported in part by the National Science Foundation under Grant No. ECCS-1711702 and the U.S. Department of Transportation through the Data-Supported Transportation Operations and Planning (D-STOP) Tier~1 University Transportation Center.}}

\begin{document}

\maketitle

\begin{abstract}
 A fully-digital wideband joint communication-radar (JCR) with a multiple-input-multiple-output (MIMO) architecture at the millimeter-wave (mmWave) band will enable high joint communication and radar performance with enhanced design flexibility.  A quantized receiver with few-bit analog-to-digital converters (ADCs) will enable a practical JCR solution with reduced power consumption for futuristic portable devices and autonomous vehicles. In this paper, we present a joint communication-radar proof-of-concept platform, named JCR70, to evaluate and demonstrate the performance of these JCR systems using real channel measurements in the 71-76 GHz band. We develop this platform by extending a mmWave communication set-up with an additional full-duplex radar receiver and by capturing the MIMO JCR channel using a moving antenna on a sliding rail. To characterize the JCR performance of our developed tested, we conduct several indoor and outdoor experiments and apply traditional as well as advanced processing algorithms on the measured data.  Additionally, we compare the performance of our JCR70 platform with the INRAS Radarbook, which is a state-of-the-art automotive radar evaluation platform at 77 GHz. The results demonstrate that a quantized receiver with 2-4 bit ADCs generally performed quite close to the high-resolution ADC for a signal-to-noise ratio of up to 5 dB. Our JCR70 platform with a fully digital JCR waveform at 73 GHz and 2 GHz bandwidth achieved higher resolution capability than the Radarbook due to higher bandwidth and larger synthesized antenna aperture.

\end{abstract}

\IEEEpeerreviewmaketitle

%%%%%%%%%%%%%%%%%%%%%%%%%%%%%%%%%%%%%%%
\section{Introduction}
%%%%%%%%%%%%%%%%%%%%%%%%%%%%%%%%%%%%%%%

Millimeter-wave (mmWave) communication and radar are key technologies for many demanding applications, such as automated driving~\cite{ChoVaGon:Millimeter-Wave-Vehicular-Communication:16} and smart connected devices~\cite{lien2016soli}. MmWave radars enable high-resolution sensing with a wide field of view, while mmWave communications provide a high data rate.  The combination of these two technologies into a single joint communication-radar (JCR) enables hardware reuse and a common signaling waveform. This leads to significant benefits in cost, power consumption, latency, spectrum efficiency, and market penetrability. 

In this paper, we present a fully-digital multiple-input-multiple-output (MIMO) JCR platform, named JCR70, at 71-76 GHz band. Due to the fully-digital MIMO functionality and a software-defined architecture, JCR70 provides enhanced communication and radar performance with increased waveform and beamforming design flexibility. Unfortunately, a naive design of a fully-digital MIMO systems at the mmWave band will result in high cost, hardware complexity, and power consumption due to large bandwidths and high dimensions. To mitigate these issues, we synthesize a fully-digital MIMO testbed by moving an antenna connected with an RF chain and a high-speed, high-resolution analog-to-digital converter (ADC) on a slider. Further improvements can be achieved by using a low quantized receiver, as proposed in our theoretical paper~\cite{KumMazMez:Low-Resolution-Sampling-for-Joint:18}. Therefore, we also perform a proof-of-concept evaluation for a fully-digital JCR with low- and medium- resolution ADCs by emulating quantization effect on the data collected from JCR70 testbed. This will enable a practical fully-digital MIMO JCR solution for futuristic radio systems.

Because of hardware limitations, mmWave JCR prototyping using communication testbeds have been difficult~\cite{AjoSreLoc:On-the-Feasibility-of-Using-IEEE:19}. Recently, \cite{Kim:Experimental-Demonstration-of-MmWave:19} investigated the applicability of the IEEE 802.11ad technology at 60 GHz for communications on a vehicular testbed using the Tensorcom 802.11ad module. In~\cite{AjoSreLoc:On-the-Feasibility-of-Using-IEEE:19}, the feasibility of IEEE 802.11ad-based radar~\cite{KumChoGon:IEEE-802.11ad-based-Radar::17,GroLopVen:Opportunistic-Radar-in-IEEE:18} was performed indoors in the range domain using a Dell laptop with IEEE 802.11ad functionality at 60 GHz. The IEEE 802.11ad testbeds developed in~\cite{Kim:Experimental-Demonstration-of-MmWave:19,AjoSreLoc:On-the-Feasibility-of-Using-IEEE:19}, however, used analog beamforming and was not fully programmable. Additionally, the strong atmospheric absorption at 60 GHz makes it difficult for future outdoor-to-indoor communications, when compared to 71-76  GHz~\cite{RapR.WMur:Millimeter-Wave-Wireless:14}. The existing mmWave testbeds with software-defined radio architecture and fully digital waveform generation/processing have demonstrated gigabits-per-second communication data rate at 71-76 GHz band~\cite{CudKovTho:Experimental-mm-wave-5G-cellular:14,GomSisRib:Will-COTS-RF-Front-Ends:18}. These mmWave communication prototypes, however, used analog processing in the angular domain and were not leveraged for simultaneous radar operations.

To realize a practical fully-digital radio system using low-resolution ADCs with a high sampling rate, there is some theoretical work on MIMO communications~\cite{LiuLuoXio:Low-Resolution-ADCs-for-Wireless:19,MezAntNos:Multiple-parameter-estimation:10,MoSchHea:Channel-Estimation-in-Broadband:18}. In~\cite{MezAntNos:Multiple-parameter-estimation:10}, an iterative channel estimation method using Expectation-Maximization (EM) was proposed for the ultra-high frequency band. To leverage the sparsity in the mmWave MIMO channels, approximate message passing-based channel estimation algorithms were proposed and numerically analyzed in~\cite{MoSchHea:Channel-Estimation-in-Broadband:18}. Prior work, though, did not consider the self-interference effect that occurs in a full-duplex radar operation~\cite{LiuLuoXio:Low-Resolution-ADCs-for-Wireless:19,MezAntNos:Multiple-parameter-estimation:10,MoSchHea:Channel-Estimation-in-Broadband:18}. There is limited work on low-resolution ADCs for a full-duplex radar that transmits and receives simultaneously. In~\cite{ZahNagMod:One-Bit-Compressive-Radar:19}, time-varying thresholds and $\ell_1$-norm minimization was proposed for a single-input-single-output radar with 1-bit ADC. To realize a fully-digital wideband mmWave MIMO JCR,~\cite{KumMazMez:Low-Resolution-Sampling-for-Joint:18,MazMezHea:Low-Resolution-Millimeter-Wave:18} proposed the use of high-speed, low-resolution ADCs receivers. The Cram\'er Rao bounds demonstrated that the radars with 1-bit ADC perform closely to the infinite-bit ADC for a single-target scenario at low signal-to-noise ratio (SNR)~\cite{KumMazMez:Low-Resolution-Sampling-for-Joint:18,MazMezHea:Low-Resolution-Millimeter-Wave:18}. A fully-digital mmWave MIMO JCR proof-of-concept platform with high-speed, low-resolution ADCs, however, is unavailable.

In this paper, we present a low-complexity mmWave proof-of-concept platform for a fully-digital MIMO JCR using high-, medium-, and low-resolution ADCs at 71-76 GHz band and 2 GHz bandwidth. To perform mmWave JCR characterization of the communication-centric JCR testbed, we conduct several experiments in a static indoor and outdoor setting with multiple scatters in the range and angle domain. The main contributions of the proposed research are summarized as follows:
\begin{itemize}
\item We present a fully-digital mmWave MIMO platform for demonstrating and evaluating the performance of a wideband JCR system. We develop this testbed by first extending the National Instruments (NI) mmWave platform for 5G communications~\cite{niTestbed,CudKovTho:Experimental-mm-wave-5G-cellular:14} to a single-input-single-output (SISO) JCR mmWave testbed with a full-duplex radar receiver. Then, we synthesize a single-input-single-output (SIMO) testbed by moving the transmit (TX) antenna element on a slider. To our knowledge, this will be the first software-defined radio prototype for wideband MIMO joint communication and radar with a fully digital waveform generation/processing at the mmWave band.

\item We conduct measurement campaigns to collect mmWave JCR data using trihedral corner reflectors for a single- and a two-target scenarios. We also perform experiments for extended target scenarios using a bike in the indoor setting and using a car in the outdoor setting.

\item To estimate the JCR channel from the collected data using our testbed, we apply traditional FFT-based linear processing as well as advanced Bernoulli Gaussian (BG)-Generalized Approximate Message Passing (GAMP) and Gaussian mixture (GM)-GAMP algorithms with sparsity constraints. Additionally, we use the EM technique to optimize the hyperparameters of the BG or GM prior. The GAMP processing provides an enhanced channel estimate with reduced sidelobes and noise at the cost of higher computational complexity than the traditional FFT-based technique.

\item We compare the radar channel estimates obtained using our fully-digital JCR70 testbed against that from a state-of-the-art automotive radar. We use the Radarbook by INRAS~\cite{inras}, which is a leading automotive radar evaluation platform for rapid prototyping at 77 GHz band with 1 GHz bandwidth. The Radarbook platform uses frequency-modulated continuous-wave (FMCW) waveforms, employs analog pre-processing in the time domain, and supports time-division multiplexing-MIMO with 12-bit ADCs. Our experimental results demonstrate that our JCR70 testbed estimates the radar channel with higher resolution than the Radarbook.

\item For evaluating the performance of low- and medium- resolution ADCs, we collect measurements at the full-duplex radar receiver using 12-bit ADCs and emulate the quantization effect on the collected data, such as 1-bit ADC can be emulated by just keeping the most significant bit. We mitigate the self-interference from the JCR transmitter to the radar receiver by using directive TX antenna as well as by separating the TX and receive (RX) antennas. To our knowledge, this is the first experimental evaluation of a wideband fully-digital mmWave MIMO JCR with high-speed, few-bit resolution ADCs.

\item We analyze the performance of emulated $b$-bit ADC data using the normalized mean square error (NMSE) metric for radar channel estimate in the single-target, two-target, and extended target scenarios. Additionally, we also compare the communication and radar performances of the joint system with normalized mean square error (NMSE) for both 1-bit ADC and with the infinite-resolution ADC for the proposed system model. Our experimental results show that low-resolution ADCs perform close to the high-resolution ADCs. The performance gap reduces with decreasing SNR and increasing channel sparsity.

\end{itemize}

This paper is a significant extension of our submitted conference papers~\cite{KumMazHea:A-MIMO-Joint-Communication-Radar:20,KumMezHea:A-Low-Resolution-ADC-Proof-of-Concept-Development:20}. In addition to the detailed exposition, we have included multiple point target and extended target results, outdoor testing, JCR performance using $b$-bit ADCs, and comparison with traditional radars to demonstrate and evaluate the performance of our testbed.

The rest of this paper is organized as follows. We formulate a system model for our proposed JCR system in Section~\ref{sec:SystemModel}. In Section~\ref{sec:HardwarePlatform}, we describe our developed fully-digital SIMO hardware testbed with 2 GHz bandwidth at 71-76 GHz band. Then, we outline the software platform for our testbed with traditional as well as advanced receive processing algorithms in Section~\ref{sec:SoftwareTestbed}. In Section~\ref{sec:Results}, we describe the experimental and numerical results. Finally, we conclude our work and provide direction for future work in Section~\ref{sec:Conclusion}.

\textbf{Notation:} We use the following notation throughout the paper: The notation $\mathcal{N}(\mu,\sigma^2)$ is used for a complex Gaussian random variable with mean $\mu$ and variance $\sigma^2$. The operator $\vert \vert \cdot \vert \vert$ represent the square norm of a vector. The notation $(\cdot)^\rmT$, $(\cdot)^*$, and $(\cdot)^\rmc$ stand for transpose, Hermitian transpose, and conjugate of a matrix/vector, while $(\cdot)^{-1}$ represent the inverse of a square full-rank matrix. Additionally, $\mathrm{vec}(\cdot)$ vectorizes a matrix to a long vector, while $\circ$ represent Khatri-Rao product of matrices. The $b$-bit quantization function $\mathcal{Q}_b(\cdot)$ is applied component-wise and separately to the real and imaginary parts.

 %%%%%%%%%%%%%%%%%%%%%%%%%%%%%%%%%%%%%%%
\section{System model} \label{sec:SystemModel}
%%%%%%%%%%%%%%%%%%%%%%%%%%%%%%%%%%%%%%%

In this section, we present the system model and assumptions pertaining to the development of radar digital signal processing (DSP) receiver algorithms for our JCR70 testbed with a fully-digital communication-centric waveform, MIMO architecture, and the use of high-speed ADCs. Although we did not incorporate some hardware impairments in the system model, like phase noise, and power amplifier non-linearity, they will be taken into account in our experimental proof-of-concept evaluation. As a first step to investigate the feasibility of a mmWave JCR system with low-resolution ADCs, we assume a static indoor setting. The proof-of-concept evaluation for dynamic scenarios is a subject of future work.

\begin{figure}[!t]
\centering
\includegraphics[clip,width=\columnwidth]{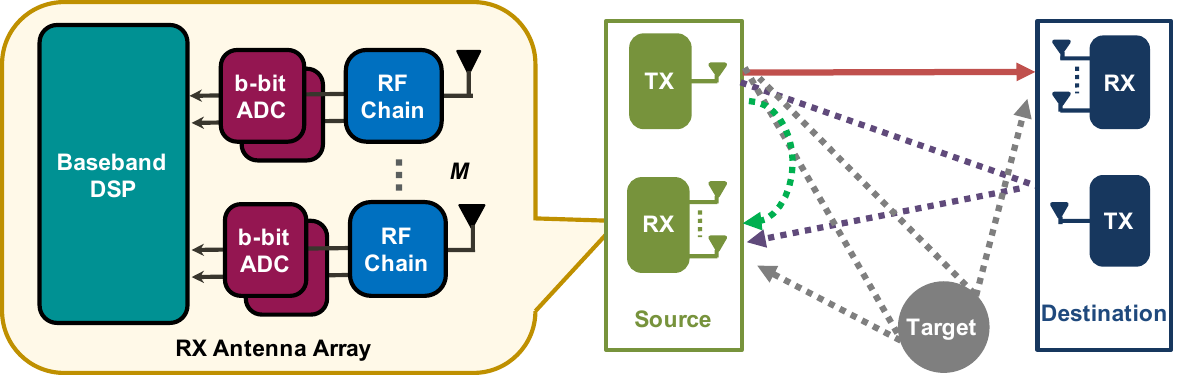}
 \vspace{-0.2cm}
 \caption{ A full-duplex joint communication-radar scenario, where first a source vehicle sends preamble to the communicating target vehicle, while simultaneously receiving its radar echoes in the presence of self-interference. Then, both vehicles start communicating data in a full-duplex mode. The JCR receivers use low-resolution ADCs per receive RF chain to reduce the hardware complexity.}
 \label{fig:systemModel}
 \vspace{-0.7cm}
\end{figure}

We consider a mmWave JCR system, where a full-duplex source transmits the JCR waveform at a carrier wavelength $\lambda$ with a signaling bandwidth $W$ to a destination receiver at a distance $d_\comm$, while simultaneously receiving echoes from the surrounding targets, as shown in Fig.~\ref{fig:systemModel}. The receiver employs an $M$ element uniform linear array (ULA) with a $b$-bit ADC per antenna with an inter-element spacing of $d_0 \leq \frac{\lambda}{2}$. Nonetheless, the proposed technique can be employed to other arrays by using $b$-bit ADC for each antenna. The TX and RX antennas are assumed to be closely separated by a distance of $d_\mI$ m to reduce the self-interference due to the full-duplex operation, while making sure that the observed range and direction of a target is the same. The single directional TX antenna is assumed to have a notch in its antenna pattern aligned with the end-fire direction, allowing further suppression of direct self-interference due to full-duplex operation. The residual direct signal leakage of the TX signal to the RX antennas is incorporated in the system model. The analysis in this paper can be extended to TX antenna array with $b$-bit digital-to-analog converters (DACs), while maintaining a separate radio-frequency chain per antenna.

In a $T$ second coherent processing interval, we consider a single carrier physical layer TX JCR waveform with $\delta$ fraction of preamble symbols and $1-\delta$ fraction of communication data symbols. We assume that the training sequences possess good correlation properties for quantized channel estimation and has a low peak-to-average power ratio~\cite{MoSchHea:Channel-Estimation-in-Broadband:18}. The mmWave WLAN standard~\cite{ieee2012wireless} with Golay complementary sequences or 5G communications~\cite{KabPat:Zadoff-Chu-Spreading-Sequence:20} using Zadoff-Chu (ZC) sequences can realize this JCR frame structure. Similar to~\cite{KumChoGon:IEEE-802.11ad-based-Radar::17}, we exploit the training sequences used in the preamble with good properties for radar sensing. The preamble is denoted by an $N$-element vector $\bt \in \mathbb{C}^{N \times 1}$. The complex baseband JCR signal model at the source transmitter with $\Ts$ symbol period is
\begin{equation} \label{eq:Tsignal}
s_{\txm}(t) = \sum_{n=-\infty}^{\infty} s[n] g_\txm(t-nT_s), \\
\end{equation}
where $s[n]$ is the transmitted symbol with $\e{\vert s[n]} \vert^2 = \Es$, and $g_\txm(t)$ is the unit energy pulse-shaping filter. The transmitted symbol $s[n]$ could either correspond to the data part or the training sequence of the JCR waveform.

We represent the mmWave communication channel using a geometric wideband channel model with $N_\mcom$ sparse clusters. Each $n^\thm$ cluster is further assumed to consist of $P_\mcom[n]$ rays/paths between the source transmitter and the destination receiver. Each $p^\thm$ ray in the $n^\thm$ cluster is characterized by its complex channel power $G_\mcom[n,p]$ that includes path loss and antenna gain, physical angle-of-arrival (AoA) relative to broadside $\phi_\mcom[n,p] \in [ -\pi/2 , \pi/2 ]$, spatial AoA $\theta_\mcom[n,p] = \frac{d_0}{\lambda} \sin(\phi_\mcom[n,p])$, and delay $\tau_\mcom[n,p]$. For the $M$-element RX ULA, the array response vector $\textbf{a}(\theta) \in \mathbb{C}^{M \times 1}$ is 
\begin{equation}
\textbf{a}(\theta) = \frac{1}{\sqrt{M}} \left[ 1, e^{-j 2 \pi \theta }, \cdots , e^{-j (M-1) 2 \pi \theta} \right]^\mathrm{T}.
\end{equation}
Denoting $G_\mrx(\theta[n,p])$ as the RX antenna element gain, $u_\mcom$ as the path loss exponent for communication channel, $L_\mcom$ as the communication loss factor that includes various losses like cable loss, impedance mismatch, the channel power $G_\comm[k,p]$ using the free space reference distance path loss model is expressed as~\cite{RapMacSam:Wideband-Millimeter-Wave-Propagation:15}
\begin{equation}
G_\mcom[n,p] =  \frac{{ \lambda^2 G_\mrx(\theta[n,p]) }}{(4 \pi)^2 (d_\mcom[n,p])^{u_\mcom} L_\mcom },
\end{equation}
where $u_\mcom$ is close to 2 for the mmWave line-of-sight (LoS) communications in outdoor urban \cite{RapMacSam:Wideband-Millimeter-Wave-Propagation:15} and rural scenarios~\cite{ MacSunRap:Millimeter-Wave-Wireless:16}. Under this model, the communication channel $\bh_\mcom(\tau) \in \mathbb{C}^{M\times 1}$ corresponding to delay $\tau$ with $g_\mrx(t)$ as the RX pulse shaping function is expressed as 
\begin{equation} \label{eq:commCh1}
\bh_\mcom(\tau) =   \sum_{n = 0}^{N_\mcom-1} \sum_{p = 0}^{P_\mcom[n]-1} \sqrt{G_\mcom[n,p]} \mathbf{a}(\theta_\mcom[n,p]) g_\mrx(\tau-\tau_\mcom[n,p]).
\end{equation}

The radar channel is assumed to consist of $N_\mr$ clusters, which includes a small-delay cluster of residual self-interference due to the simultaneous TX and RX operation in radars, along with the other $N_\mr-1$ clusters corresponding to the reflections from multiple surrounding objects. The maximum delay spread of the radar channel is assumed to be $\tau_{\mathrm{max}}$, which is generally much smaller than the communication delay spread. Each of the $p^\thm$ ray in the $n^\thm$ cluster is characterized by its complex channel power $G_\mrad[n,p]$ that includes path loss, antenna gain, and radar cross-section (RCS) $\sigma_{\mathrm{RCS}}[n,p]$, physical angle-of-arrival (AoA) $\phi_\mrad[n,p]$, spatial AoA $\theta_\mrad[n,p] = \frac{d_0}{\lambda} \sin(\phi_\mrad[n,p]) $, and delay $\tau_\mrad[n,p]$.  The radar targets in this paper are either single/multiple point reflectors or extended targets. The two-way radar channel power, $G_\mr[n,p]$, corresponding to the $p^{\mathrm{th}}$ path in the $n^\thm$ LoS cluster with $L_\mrad$ as the radar loss factor, which includes various losses like cable loss, is given as
\begin{equation}
G_\mr[n,p] = \frac {\lambda^2  G_\mrx(\theta[n,p]) \sigma_{\mathrm{RCS}}[n,p]}{64 \pi^3 (d_\mr[n,p])^{u_\mr} L_\mrad} .
\end{equation}
We will experimentally estimate the path loss exponent $u_\mr$ for mmWave radar channels at 73 GHz using our set-up. We choose the cluster with $n = 0$ to represent the residual self-interference effect. Under this model, the radar channel with residual self-interference $\bh_\mI(\tau)$ is given as
\begin{equation} \label{eq:radCh1}
\begin{aligned}
\bh_{\mr}(\tau) &= \sum_{n = 1}^{N_\mr-1} \sum_{p = 0}^{P_\mr[n]-1} \!\! \sqrt{G_\mr[n,p]} \mathbf{a}(\theta_\mr[n,p]) g_\mrx(\tau-\tau_\mr[n,p])  \\&+ \bh_\mI(\tau),
\end{aligned}
\end{equation}
where $\bh_\mI(\tau) \! = \! \sum_{p = 0}^{P_\mr[0]-1} \!\!  \sqrt{G_\mr[0,p]} \mathbf{a}(\theta_\mr[0,p]) g_\mrx(\tau-\tau_\mr[0,p])$.

While the physical radar channel vector in~\eqref{eq:radCh1} is accurate, it is difficult to estimate because of the non-linear dependence on the unknown parameters, such as channel powers, delays, and AoAs. Due to the finite waveform bandwidth, however, the radar channel model can be represented by a discretized channel vector $\bh_\rmd[k]$ by uniformly sampling the delay domain at the Nyquist rate $\Delta\tau = 1/W$ with the aid of Fourier series expansion. Denoting $K = \lceil W \tau_{\mathrm{max}} + 1 \rceil $ range bins of delay resolution  $\Delta\tau = 1/W$, the frequency domain representation $\tilde{\bh}_\mr(f)$ of $\bh_\mr(\tau)$ is given by~\cite{bajwa2010compressed}
\begin{equation} \label{eq:radCh2}
\tilde{\bh}_\mr(f)  =    \sum_{k = 0}^{K-1}  \bh_\rmd[k] e^{j2 \pi \frac{k} {W }f}.
\end{equation}
Similarly, due to the finite RX antenna aperture, the channel at the $k^\thm$ range bin can be represented by a discretized channel vector $\bx_k$ with $M$ discrete angle bins of spatial angle resolution $\Delta\theta = 1/M$, such that they are related using the Fourier series expansion in the angle domain, and is given by
\begin{equation} \label{eq:radCh3}
\bh_\rmd[k] =  \sum_{m=0}^{M-1} x_k[m] \mathbf{a} \left( \frac{m}{M} \right) = \bA_\rmM \bx_k,
\end{equation}
where $\mathbf{a} \left( m/M \right)$ is the $m^\thm$ column of the RX beamforming matrix $\bA_\rmM \in \mathbb{C}^{M \times M}$.

Using the discretized channel matrix $\bX_\mr = [ \bx_0 \text{ } \cdots \text{ } \bx_{K-1}]$ in the range-angle domain, we can express the unquantized RX signal model in a linear equation format. Therefore, the unquantized RX radar signal corresponding to the training sequence vector $\bt $ with $\bD \in \mathbb{C}^{K \times N}$ as the circulant-shift matrix of $\bt$, where the $k^\thm$ row of $\bD$ is obtained by circularly shifting $\bt^\mathrm{T}$ by $k$, and the additive Gaussian noise matrix $\bW_\mr \in \mathbb{C}^{M \times N}$ with zero mean and variance $\sigma_{w}^2$ is given as
\begin{equation} \label{eq:rady}
\bY_{\mr} =     \bA_\rmM \bX_\mr  \bD+ \bW_{\mr} .
\end{equation}
We define the radar SNR as $\e{\vert \vert \bz_\mr \vert \vert^2}/\e{\vert \vert \bw_\mr \vert \vert^2}$ with the unquantized noiseless signal vector $\bz_\mr \triangleq \mathrm{vec}(\bA_\rmM \bX_\mr  \bD)$ and the noise vector $\bw_\mr \triangleq \mathrm{vec}(\bW_\mr)$. 

The quantized RX radar signal using $b$-bit ADC is $\bQ_\mr =\mathcal{Q}_b\Big(  \bY_{\mr}\Big)$. The quantization operation introduces distortion due to the granularity of the quantizer and due to the clipping effects. These distortions cause the quantized RX signal to be non-linearly related to the radar channel matrix $\bX_\mr$ unlike the unquantized case.
Denoting the discretized radar channel vector as $\bx_\mr \triangleq \mathrm{vec}(\bX_\mr)$, the quantized complex-baseband RX radar signal vector defined as  $\bq_\mr \triangleq \mathrm{vec}(\bQ_\mr)$ is represented as
\begin{equation}\label{eq:qRad}
\bq_\mr =  \mathcal{Q}_b\Big( (\bD^\rmT \kron \bA_\rmM ) \bx_\mr + \bw_\mr \Big).
\end{equation}
The unquantized RX communication signal $\bq_\mcom$ can similarly be expressed using the discretized communication channel vector $\bx_\mcom$  in the range-angle domain as in~\eqref{eq:qRad}. To estimate the discretized JCR channel matrix in the range-angle domain using the quantized received JCR signal, we can either use the traditional method with a correlation-based time-domain processing algorithm and a fast Fourier transform (FFT)-based angle-domain technique~\cite{KumChoGon:IEEE-802.11ad-based-Radar::17}, or using advanced GAMP algorithms. These algorithms are described in detail in Section~\ref{sec:SoftwareTestbed}. 

For radar performance evaluation,  we consider 1-bit to 8-bit ADC and compare it with 12-bit ADC available in our JCR testbed. The main advantage of the low-resolution architecture is that it can be implemented with low power consumption and reduced the overall complexity of the circuit~\cite{SinPonMad:Multi-Gigabit-communication:-the-ADC-bottleneck1:09}.

 %%%%%%%%%%%%%%%%%%%%%%%%%%%%%%%%%%%%%%%
\section{Experimental JCR70 hardware platform} \label{sec:HardwarePlatform}
%%%%%%%%%%%%%%%%%%%%%%%%%%%%%%%%%%%%%%%
%XXX Check CAR-STOP report for more details XXX

\begin{figure}[!h]
\centering
\includegraphics[width=0.55\columnwidth]{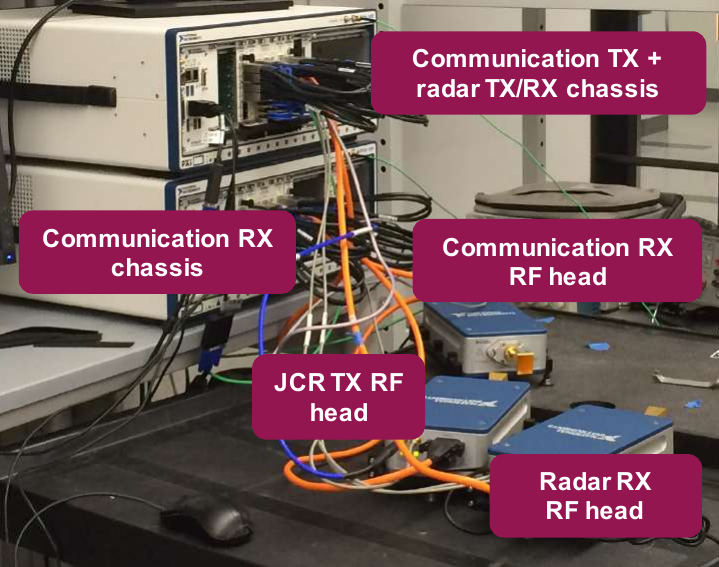}
 %\vspace{-1.2em}
 \caption{The mmWave joint communication and radar set-up with mono-static radar and bi-static communication in a SISO configuration. The radar and communication share a common fully-digital waveform to enable hardware/spectrum reuse. The radar is in a full-duplex mode and the interference between the radar TX and RX depends on the distance between them or the isolation provided by the objects between them.}
\label{fig:JCRTestbed}
\end{figure}

\begin{figure*}[!h]
\centering
\includegraphics[width=1.6\columnwidth]{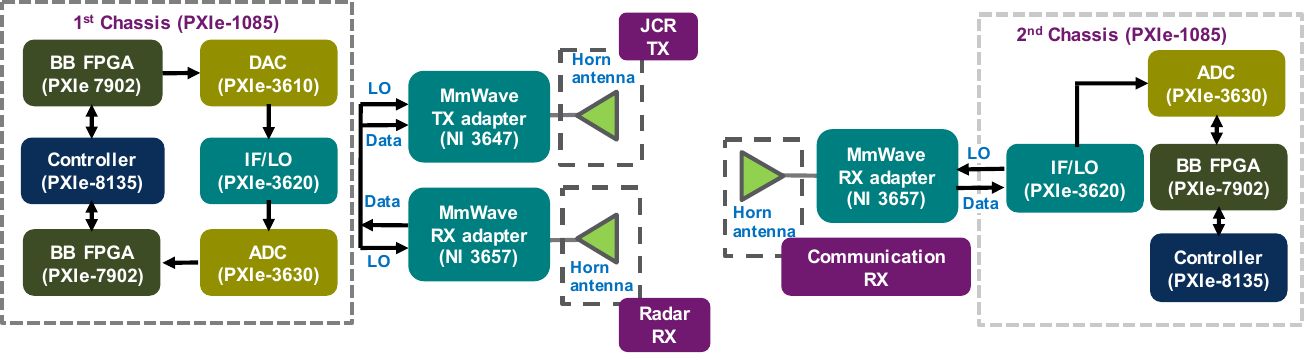}
 %\vspace{-1.2cm}
\caption{The block diagram of a SISO joint communication and radar testbed. The PXIe numbers correspond to parts from NI.}
\label{fig:JCRBlock}
\vspace{-1.0em}
\end{figure*}
This section describes the hardware for our mmWave wideband JCR70 platform. Our JCR70 platform is developed for the use case where a source JCR transmitter sends a signal to a communication RX and uses the echoes from surrounding targets and clutter to derive target range and AoA estimates at the source radar receiver.  First, we developed a full-duplex SISO JCR set-up with one JCR transmitter, one communication receiver, and one radar receiver. Then, we extended this set-up for a SIMO mmWave JCR system by moving TX antenna on a slider using a stepper motor to collect RX signals with multiple TX-RX inter-spacing simultaneously for communication and radar receivers. %For evaluating the performance of our JCR testbed, we used corner reflectors for precise characterization and extended targets such as a bike, and a car for JCR characterization in a more realistic setting. 

\subsection{SISO JCR testbed}
The mmWave JCR testbed in a SISO set-up with 2~GHz bandwidth is shown in Fig.~\ref{fig:JCRTestbed}.  This set-up extends the mmWave communication testbed developed by NI~\cite{niTestbed} for JCR functionality in a full-duplex configuration. We developed the JCR testbed using two NI PXIe-1085 express chassis. One of the chassis acts as the source JCR that consists of a communication transmitter and radar receiver and the other chassis acts as the destination receiver for the communication receiver. Each chassis houses NI PXIe-8135 controller, NI PXIe 7902 FPGA for baseband TX/RX processing, NI PXIe 3610 DAC module, NI PXIe 3630 ADC module, and NI PXIe 3630 for intermediate frequency (IF) up-/down-conversion. The IF-local oscillator (LO) module is connected to mmWave TX/RX head(s) for up-/down-conversion to 71-76 GHz band and then these mmWave heads are connected to the horn antennas for over-the-air JCR transmission. The two chassis can be synchronized using Rubidium clock. The block diagram of this set-up is shown in Fig.~\ref{fig:JCRBlock} and the specifications for these modules are given in Table~\ref{tab:HardSpecs}.

\begin{table*}[h]
\caption{Hardware specifications for our JCR70 testbed}
\centering
%     \begin{adjustbox}{width=\textwidth,center}
    % \begin{adjustbox}{center}
        \begin{tabular}{l|ll}
        \hline 
        \textbf{Hardware} & \textbf{Description/Specifications} & \textbf{Value(s)} \\\hline
            \hline
            NI PXIe-7902 Baseband & FPGA  & Virtex-7 485T \\\hline
           \multirow{3}{*}{NI PXIe 3610 DAC} & \multicolumn{1}{l}{Resolution} & \multicolumn{1}{l}{14 bit} \\\cline{2-3}
                                 & \multicolumn{1}{l}{Sampling rate} & \multicolumn{1}{l}{3.072GS/s} \\\cline{2-3}
                                 & \multicolumn{1}{l}{Bandwidth} & \multicolumn{1}{l}{2 GHz} \\\hline
             NI PXIe 3620 IF-LO & IF tuning range & 8.5 -13.5 GHz \\\hline                    
           \multirow{3}{*}{NI PXIe 3630 ADC} & \multicolumn{1}{l}{Resolution} & \multicolumn{1}{l}{12 bit} \\\cline{2-3}
                                 & \multicolumn{1}{l}{Sampling rate} & \multicolumn{1}{l}{3.072GS/s} \\\cline{2-3}
                                 & \multicolumn{1}{l}{Bandwidth} & \multicolumn{1}{l}{2 GHz} \\\hline
            \multirow{2}{*}{NI 3647 RX mmWave head} & \multicolumn{1}{l}{Frequency band} & \multicolumn{1}{l}{71-76 GHz} \\\cline{2-3}
                                 & \multicolumn{1}{l}{Quantities} & \multicolumn{1}{l}{2} \\\hline
              \multirow{2}{*}{NI 3657 TX mmWave head} & \multicolumn{1}{l}{Frequency band} & \multicolumn{1}{l}{71-76 GHz} \\\cline{2-3}
                                 & \multicolumn{1}{l}{Quantities} & \multicolumn{1}{l}{1} \\\hline
              \multirow{2}{*}{Horn antennas} & \multicolumn{1}{l}{Gains} & \multicolumn{1}{l}{23 dBi and 10 dBi} \\\cline{2-3}
                                 & \multicolumn{1}{l}{Beamwidth} & \multicolumn{1}{l}{10 and 50 degrees} \\\hline
\multirow{2}{*}{Trihedral corner reflectors antennas} & \multicolumn{1}{l}{Edge length} & \multicolumn{1}{l}{4.3 and 3.2 inches} \\\cline{2-3}
                                 & \multicolumn{1}{l}{Radar cross-section} & \multicolumn{1}{l}{5 and 8 dBsm at 73 GHz} \\\hline
  %   Integrated stepper drive and motor device &  Microstep resolution & 200 to 51, 2000 steps/rev step  \\\hline    
  \multirow{2}{*}{ Sliding rail} & \multicolumn{1}{l}{Length} & \multicolumn{1}{l}{21 cm} \\\cline{2-3}
                                 & \multicolumn{1}{l}{Synthetic antenna inter-spacing} & \multicolumn{1}{l}{1.69 mm} \\\hline
                                 %& \multicolumn{1}{l}{Maximum number of synthetic antennas} & \multicolumn{1}{l}{100} \\\hline                                         
             %Sliding rail &  Length &   21 cm \\                \hline
   
% & 1.69 mm \\
        \end{tabular}
%     \end{adjustbox}
%     \vspace{ - 05 mm}
    
    \label{tab:HardSpecs}
\end{table*}

This test-bed can be used for both real-time JCR prototype mode and real-time JCR channel sounding mode. To characterize the signal processing performance of the wideband mmWave JCR, we explore the JCR channel sounding mode that is more flexible and easy to use as compared to the real-time JCR prototype. The JCR channel sounding mode transmits repetitions of 2048 length ZC sequences and acquires the raw signals at both radar and communication receivers simultaneously. We then further process this acquired RX signals in MATLAB for JCR performance evaluations as described in Section~\ref{sec:SoftwareTestbed}. 

%%%%%
\subsection{SIMO JCR testbed}
 \begin{figure}[!h]
\centering
\includegraphics[width=0.7\columnwidth]{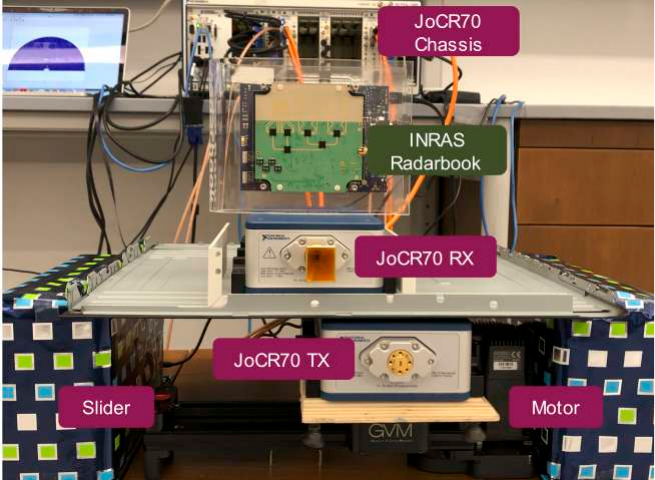}
 \caption{The SIMO JCR70 hardware platform, where the sliding motor is used to synthesize multiple digital RF chains.}
\label{fig:TXSlider}
\end{figure}

We have also synthesized a SIMO testbed by moving TX antenna on a slider using a stepper motor to collect RX signals with multiple TX-RX inter-spacing for communication and radar receivers simultaneously, as shown in Fig.~\ref{fig:TXSlider}. The inter-distance between two TX locations is kept less than or equal to half of the carrier wavelength to avoid any grating lobes.  The number of locations of the TX on the slider dictates the aperture length of the synthetic antenna, its angular resolution,  and the far-field distance. Due to the time-domain channel reciprocity,  the channel obtained using multiple TX locations and a fixed RX location will equivalently represent the channel that could have been obtained using multiple RX antenna locations with a fixed TX antenna location.  Therefore, moving the JCR transmitter to several locations with fixed RX antennas for radar and communication, we equivalently obtain the SIMO channel for both radar and communication simultaneously. In this paper, we use our mmWave SIMO JCR testbed for static joint communication and radar testing. The proof-of-concept development for dynamic scenarios is a subject of future work.

We benchmark our proposed JCR system against a state-of-the-art automotive radar. We use the Radarbook by INRAS~\cite{inras}, which is a leading automotive radar evaluation platform for rapid prototyping at 77 GHz band with 1 GHz bandwidth using an Infineon chipset. We mount the Radarbook on top of the NI radar RX adapter module, as shown in Fig.~\ref{fig:TXSlider}. The Radarbook uses FMCW waveforms, includes 4 TX and 8 RX antennas, and supports time-division multiplexing (TDM)-MIMO with 12-bit ADCs and maximum sampling rate of 80 MSps. It employs analog pre-processing in the time-domain to perform deramping that reduces the ADC sampling requirement. In the Radarbook, the software support is provided for basic functionality to control front end using MATLAB along with the direct access to the raw deramped complex radar outputs. The accessibility of the raw radar data is a main advantage of using the Radarbook versus a commercial radar that only provides access to the final estimated radar parameters~\cite{delphi}.

In the Radarbook, a virtual ULA array with $M = 29$ virtual elements and 1.948~mm element spacing is synthesized by using the TDM-MIMO technique described in~\cite{inras}. After the deramping in the analog domain and the virtual ULA construction in the digital domain, the radar output can be mathematically expressed similar to \eqref{eq:qRad} with $\bD$ as a $K$-point FFT matrix. The RX processing techniques to estimate the radar channel using the raw deramped radar output from the 29 virtual antenna elements is described in Section~\ref{sec:SoftwareTestbed}.

%%%%%%%%%%%%%%%%%%%%%%

 %%%%%%%%%%%%%%%%%%%%%%%%%%%%%%%%%%%%%%%
 \section{Experimental JCR70 software platform} \label{sec:SoftwareTestbed}
\begin{figure*}[tbp]
\centering
\includegraphics[width=1.4\columnwidth]{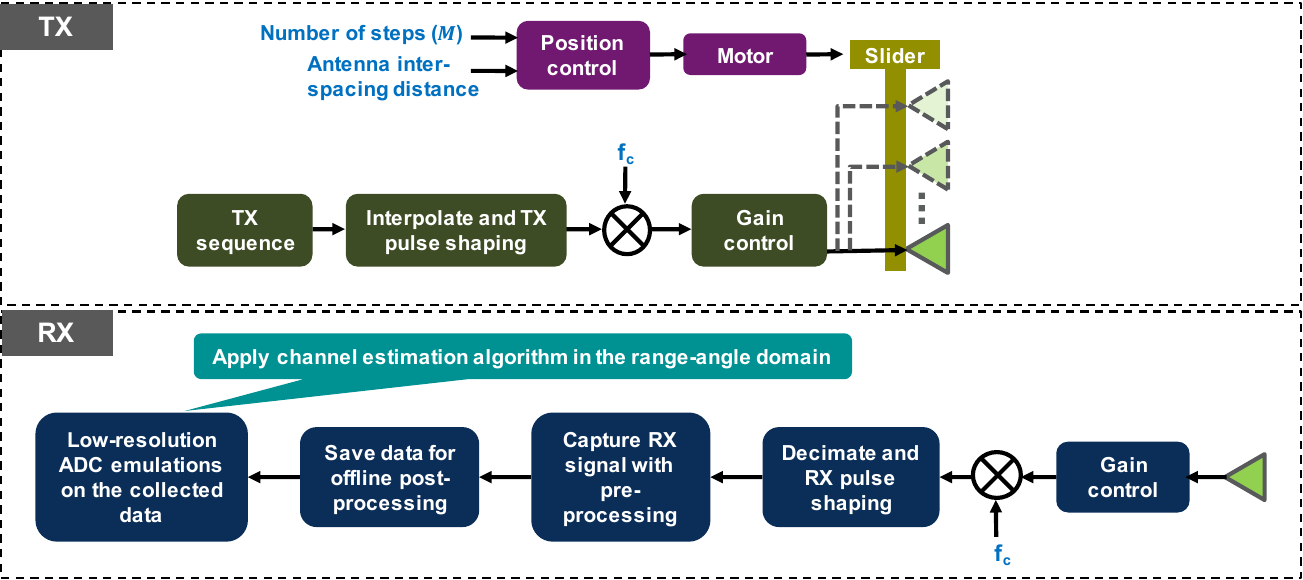}
%\vspace{-2em}
 \caption{Software block diagram for our testbed.}
\label{fig:SoftwareTestbed}
%\vspace{-1.2em}
\end{figure*}

%%%%%%%%%%%%%%%%%%%%%%%%%%%%%%%%%%%%%%% 
In this section, we outline the software platform for our JCR70 testbed, as shown in Fig.~\ref{fig:SoftwareTestbed}. First, we will describe the real-time TX signal generation and offline RX emulations for $b$-bit ADCs in presence of the self-interference effects using our JCR70 testbed.  Then, we will describe the real-time and offline RX processing algorithms. The real-time algorithms are implemented in LabVIEW and LabVIEW FPGA, whereas the offline emulations and processing are implemented in MATLAB. Additionally, we will explain the RX processing techniques for the Radarbook.

\subsection{Transmit and receive signals}
Zadoff-Chu training sequences of length 2048 are used as the training sequence. Multiple ZC sequences are sent in each scan, several scans are carried out at each TX antenna location, and the transmitter is moved on the slider for multiple steps using a motor to synthesize our SIMO testbed. For each scan at a particular TX location, the transmitter sends several repetitions of the training sequences and then wait to RX echoes for a predefined time interval of around 1~second, which is large enough to avoid any target range ambiguity. The JCR70 testbed employs around 10~seconds wait time before moving to the next TX location.

To emulate the received signal using medium- and low-resolution ADCs on the collected data, JCR70 performs uniform mid-rise quantization. For a scalar complex-valued $y$, the received quantized signal $q = \mathcal{Q}_b(y)$ is defined as
\begin{equation} \label{eq:Quantb} \begin{split}
q = \mathrm{sign}(\mathrm{Re}(x)) \left( \mathrm{min} \left(  \Bigg\lceil \frac{\mathrm{Re}(x) }{\Delta_\mathrm{Re}} \Bigg\rceil , 2^{b-1} \right) - \frac{1}{2} \right)\Delta_\mathrm{Re} + \\ \rmj  \text{ } \mathrm{sign}(\mathrm{Im}(x)) \left( \mathrm{min} \left(  \Bigg\lceil \frac{\mathrm{Im}(x) }{\Delta_\mathrm{Im}} \Bigg\rceil , 2^{b-1} \right) - \frac{1}{2} \right)\Delta_\mathrm{Im} \end{split},
\end{equation}
where $\Delta_\mathrm{Re} = \left( \e{\vert \mathrm{Re}(x)\vert^2}\right)^{0.5} \Delta_b$ and $\Delta_\mathrm{Im} = \left( \e{\vert \mathrm{Im}(x)\vert^2} \right)^{0.5} \Delta_b$. The quantization stepsize $\Delta_b$ is chosen to minimize the quantization distortion mean square error assuming a Gaussian input. The values of $\Delta_b$ can be found in~\cite{Max:Quantizing-for-minimum-distortion:60,Jay:Digital-Coding-of-Waveforms:84}.

%%%%%%%%%%%%%%%%%
\subsection{Receive processing}\label{sec:RXProc}
%%%%%%%%%%%%%%%%%%%
The raw communication or radar signal $\bq_\mcom$ or $\bq_\mr$ in \eqref{eq:qRad} received from the real-time JCR sounding testbed is used for estimating the JCR channel $\bx_\mcom$ or $\bx_\mr$ in the range-angle domain. We benchmark the radar channel estimate obtained using JCR70 against that from the Radarbook. The radar output obtained from the JCR70 or the Radarbook is processed either by using a traditional FFT-based linear processing or by using an advanced non-linear GAMP technique.  
 
In the traditional FFT-based linear processing, first the range processing is performed for each antenna location and then the FFT is applied in the angle domain for each range bin to estimate the channel in the range-angle domain. In our JCR70 testbed, the range processing involves matched filtering of the raw received data with the known training sequence for each antenna location~\cite{KumChoGon:IEEE-802.11ad-based-Radar::17}. In the Radarbook, the range processing involves applying FFT on the raw deramped output obtained from each virtual MIMO antenna element.

%\subsection{Advanced GAMP processing}
%%%%%%%%%%%%%%%%%%%%%%%%%% %%%%%%%%%%%%%
\begin{algorithm}[tbp]
\caption{EM-GM-GAMP algorithm}
\label{GAMP_alg}
\begin{algorithmic}[1]
\STATE \textbf{Input:}  Observation vector $\bq$, $\bB= \bD^\rmT \kron \bA_\rmM $ 
\STATE \textbf{Initialize:} $\bz^0=\boldsymbol{0}$,
$\bx^0=\boldsymbol{0}$, $\boldsymbol{\psi}^0={\mathrm{
const}}$, $\ell \leftarrow 0$   \\    
\REPEAT
\STATE $\ell \! \leftarrow \!  \ell+1$
\STATE Output Step:\\
 %$\boldsymbol{u}^{\ell} \! \leftarrow \! \bB\hat{\bx}^{\ell-1}  + \boldsymbol{\psi}^{\ell-1} \circ\bz^{\ell-1}$ \\
 $\bz^{\ell} \! \leftarrow \! g_\ell(-\bB\hat{\bx}^{\ell-1}  + \boldsymbol{\psi}^{\ell-1} \circ\bz^{\ell-1},\bq,\boldsymbol{\psi}^{\ell-1})$\\
$\boldsymbol{\xi}^{\ell} \! \leftarrow \! (\bB \circ \bB^*)^{\mathrm{T}}
g'_\ell ( -\bB\hat{\bx}^{\ell-1}  + \boldsymbol{\psi}^{\ell-1} \circ\bz^{\ell-1}, \bq, \boldsymbol{\psi}^{\ell-1}) $ \\
\STATE Input Step: \\
$\bx^{\ell}  \! \leftarrow \!  f_\ell(\bB^{\mathrm {H}}
\bz^\ell  + \boldsymbol{\xi}^\ell \circ
\hat{\bx}^{\ell-1},  \boldsymbol{\xi}^\ell)$ \\
$\boldsymbol{\psi}^\ell \! \leftarrow \!   (\bB \circ \bB^*) 
f'_\ell(\bB^{\mathrm{H}} \bz^\ell  +
\boldsymbol{\xi}^\ell \circ \hat{\bx}^{\ell-1},  \boldsymbol{\xi}^\ell )$
\STATE update the parameters $\boldsymbol{\Omega}$ using EM algorithm
\UNTIL{the cost does not significantly decrease or a maximum iteration count has been reached}
%\vspace{-1.6em}
\RETURN $\bx^\ell$.
\end{algorithmic}
\end{algorithm}

%%%%%%%%%%%%%%%%%%%%%%%%%%%%%%
We also perform advanced sparse reconstruction using the high-performing non-linear GAMP technique in Algorithm~\ref{GAMP_alg} to estimate the mmWave channel in the range-angle domain from the received signal obtained using our JCR70 testbed or the Radarbook~\cite{Ran:Generalized-approximate-message:11}. To leverage sparsity in the mmWave channels, we assume the channel coefficients $x_i$ of the mmWave channel in the range-angle domain, $\bx_\mr$ or $\bx_\mcom$, are drawn from the BG or GM model with unknown parameters $\boldsymbol{\Omega}$, having marginal probability distribution function
\begin{equation}
p_X(x_i;\boldsymbol{\Omega}) = \eta_0\delta(x_i)+\sum_{i=0}^{V-1} \eta_i \mathcal{N}({x_i; \mu_i,\nu_i}),
\end{equation}
where $\eta_0$ is the probability of $x_i = 0$ thereby enforcing sparsity, $\delta(\cdot)$ is the Dirac distribution, and the unknown parameters $\boldsymbol{\Omega}= [\{ \eta_i, \mu_i,\nu_i\}_{i=0}^{V-1}, V]$ with $\eta_i$ as the weight, $\mu_i$ as the mean and $\nu_i$ as the variance of the Gaussian mixture with $V$ components. By definition, $\sum_{i=0}^{V-1} \eta_i = 1$. For the BG model, $V=1$ and we use $\eta$ to represent $\eta_1$ for simplicity with $\eta_0 = (1-\eta)$.

The GAMP algorithm generally performs channel estimation better than the traditional processing in sparse environments. The GAMP algorithm, however, has more computational complexity than the traditional FFT-based linear processing. Therefore, a two-stage processing with traditional processing on the collected data with the second stage of GAMP processing on a smaller part of the channel estimate to improve the resolution would improve the resolution without increasing the complexity much.

In Algorithm~\ref{GAMP_alg}, we provide the pseudo-code of GAMP algorithm \cite{Ran:Generalized-approximate-message:11} to estimate the channel estimate $\bx_\mr$ or $\bx_\mcom$ with noiseless received signal $\bz_\mr$ or $\bz_\mcom$ based on the quantized observation $\bq_\mr$ or $\bq_\mcom$ in~\eqref{eq:Quantb}. The recursive approach breaks apart the entire estimation problem into smaller scalar estimations described by the input denoising function $f_\ell(v,\xi^\ell)$ and the output (residual) function $g_\ell(-u,q,\psi^{\ell-1})$. The minimum mean square error (MMSE) denoiser function $f_\ell(v,\xi^\ell)$ for estimating $x$ using BG-GAMP is calculated based on the posterior mean obtained from the Bernoulli-Gaussian prior $x \sim BG\left(\eta,0, \nu \right)$ along with the noisy observation $v|x\sim \mathcal{N}\left( \xi^{\ell} x, \xi^\ell \right)$:
\begin{equation}
f_\ell(v,\xi^\ell)=\e{x |v }= \zeta \frac{\nu }{\xi^\ell \nu+1} v,
\end{equation}
where
\begin{equation}
\begin{aligned}
\zeta &=\frac{\eta \mathcal{N}\left(v/\xi^\ell ;0, \nu + 1/\xi^\ell\right)}{(1-\eta) \mathcal{N}\left(v/\xi^\ell;0, 1/\xi^\ell \right)+\eta \mathcal{N}\left(v/\xi^\ell;0,\nu+1/\xi^\ell \right)}.
\end{aligned}
\end{equation}
The output (residual) function (applied element-wise for each real/imaginary component)~\cite{MezNos:Efficient-reconstruction-of-sparse:12} is
\begin{equation}
\begin{aligned}
&  g_\ell(-u,q,\psi^{\ell-1})= - \frac{u}{\psi^{\ell-1}} +\\
& ~ \frac{ {\mathrm{exp}}\left(-\frac{( q^{\mathrm {lo}}-u)^2}{\sigma_w^2+\psi^{\ell-1}}\right) - {\mathrm {exp}}\left(-\frac{( q^{\mathrm{ up}}-u)^2}{\sigma_w^2+\psi^{\ell-1}}\right)  }{2 \sqrt{ \pi (\sigma_w^2 + \psi^{\ell-1}) } \left(\mathrm{erf} \left(\frac{q^{\mathrm {up}}-u}{\sqrt{\sigma_w^2+\psi^{\ell-1}}}\right) \!- \! \mathrm{erf}\left( \frac{q^{\mathrm{low}}-u}{\sqrt{\sigma_w^2+\psi^{\ell-1}}}  \right)   \right) } \! ,
\end{aligned}
\end{equation}
with $q^{\mathrm {lo}}$ and $q^{\mathrm {up}}$ being the lower and upper quantization boundary. 
The scalar input functions $f_\ell$ and output function $g_\ell$ are applied element-wise to vectors in the GAMP algorithm. The messages exchanged between the input and output steps consist of the results of the individual scalar estimations as well as the curvature around these optima. This is crucial for faster convergence. The curvature message vectors $\boldsymbol{\xi}^{\ell}$ and $\boldsymbol{\psi}^{\ell}$ are obtained by means of the Wirtinger derivatives $g'_\ell$ and $f'_\ell$ in step (5) and (6) with respect to the first argument. Similar expressions for the GM-GAMP algorithm can be calculated, and is given in~\cite{VilSch:Expectation-Maximization-Gaussian-Mixture-Approximate:13}. Additionally, we can use the EM technique to optimize the hyperparameters of the BG or GM prior and the resulting algorithm is called EM-BG-GAMP and EM-GM-GAMP~\cite{VilSch:Expectation-Maximization-Gaussian-Mixture-Approximate:13}.

 %%%%%%%%%%%%%%%%%%%%%%%%%%%%%%%%%%%%%%%
 \section{Experimental results} \label{sec:Results}
  %%%%%%%%%%%%%%%%%%%%%%%%%%%%%%%%%%%%%%%
 In this section, we describe different experiments conducted using our mmWave wideband testbed for the proof-of-concept performance evaluation of joint communication and radar at 73 GHz. The training sequence used in our mmWave JCR testbed is ZC sequence of length $N = 2048$ and the symbol rate is 1.536 GHz.  First, we describe the hardware characterization for JCR that includes RF hardware calibration. Then, we evaluate the performance of our testbed for radar channel estimation using fully digital radar with infinite-bit and 1-bit ADC at RX. We conducted our experiments using corner reflectors for precise characterization and extended targets such as a bike and a car for JCR characterization in a more realistic setting. Finally, we present the results for the wideband joint communication and radar at mmWave band. We also compare our results using the state-of-the-art INRAS Radarbook that uses FMCW waveform at 77 GHz.

To evaluate the JCR channel estimation performance using b-bit ADCs, we use the NMSE metric that we define as
\begin{equation}~\label{eq:NMSE}
\mathrm{NMSE}(\hat{\bx}) \triangleq \e{\frac{\vert \vert \bar{\bx} - \hat{\bx} \vert \vert^2}{\vert \vert \bar{\bx} \vert \vert^2}},
\end{equation}
where $\hat{\bx}$ is the estimated channel using our emulated data for $b$-bit ADCs with $1 \leq b \leq 8$ and $\bar{\bx}$ is the channel estimate collected using our testbed at a higher SNR with 12-bit ADC and traditional FFT processing. The NMSE metric defined in \eqref{eq:NMSE} is not the traditional NMSE with $\bar{\bx}$ denoting as the true value, which is unknown. The NMSE metric in \eqref{eq:NMSE}, however,  provides relative performance evaluation of low- and medium-resolution ADCs as compared to the high-resolution ADCs.

    %%%%%%%%%%%  
\subsection{Radar: Single-target scenarios}

    %%%%%%%%%%%
  \subsubsection{Path loss}
 
\begin{figure}[htb]
\begin{minipage}[b]{\linewidth}
  \centering
  \centerline{ \includegraphics[clip,width=0.45\columnwidth]{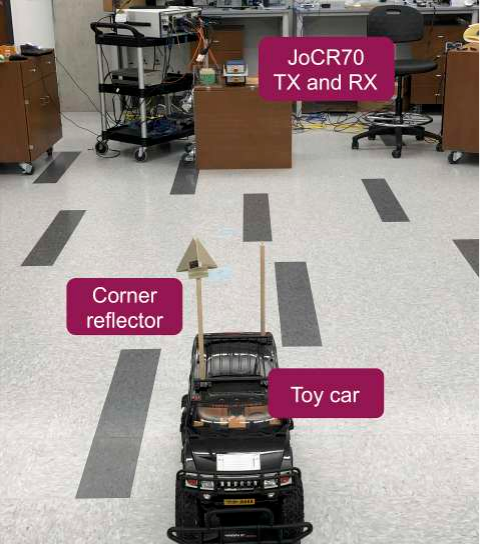}}
 % \centerline{ \includegraphics[clip,width=0.65\columnwidth]{Fig/PathlossSetup_r1}}%May13data_3
%  \vspace{1.5cm}
  \centerline{(a) Experimental set-up}\medskip
\end{minipage}
%\hspace{0.05cm}
%\hfill
\begin{minipage}[b]{\linewidth}
  \centering
  \centerline{\includegraphics[clip,width=0.69\columnwidth]{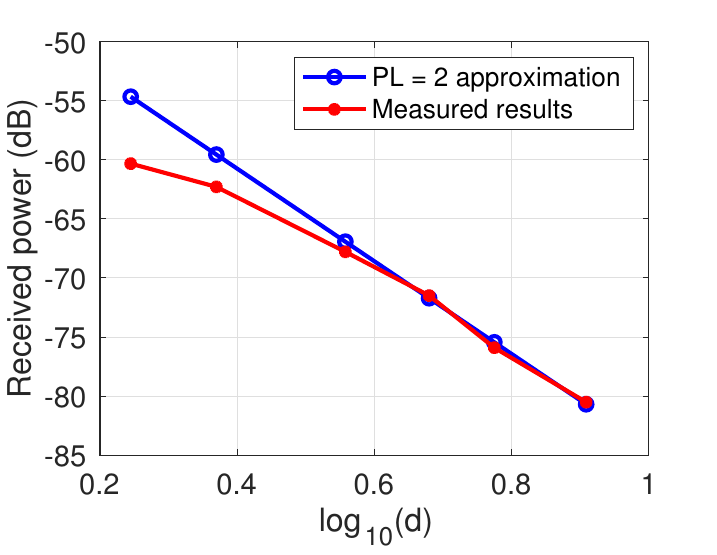}}
%  \vspace{1.5cm}
  \centerline{(b) Path loss}\medskip
\end{minipage}
\vspace{-2em}
\caption{Radar testing using JCR70 platform and a 4.3 inch trihedral corner reflector mounted on a toy vehicle for path loss measurements at different distances. The received power decrease with the distance (in meters) closely follows a path loss of 2 in the far-field region. }
\label{fig:Pathloss}
%\vspace{-1.2em}
\end{figure}
  A screenshot of the measurement campaign for indoor radar sounding using the mmWave communication channel sounder setup is shown in Fig.~\ref{fig:Pathloss}(a). Indoor data collection has been performed using this setup for different distances using corner reflectors mounted on a mobile remote-controlled 1/6 hummer car. The processed measurement results in Fig.~\ref{fig:Pathloss}(b) shows the received power of the 4.3 inch target corner reflector at different distances $d$ in meters.  With increasing distance, the amplitude of the target corner reflector decreases and the slope follows a path loss of 2 at higher distances, where it is considered in the far-field region.

  %%%%%%%%%%%%%%%%
\subsubsection{RF hardware frequency response calibration}
\begin{figure}[htb]
\begin{minipage}[b]{\linewidth}
  \centering
  \centerline{ \includegraphics[clip,width=0.7\columnwidth]{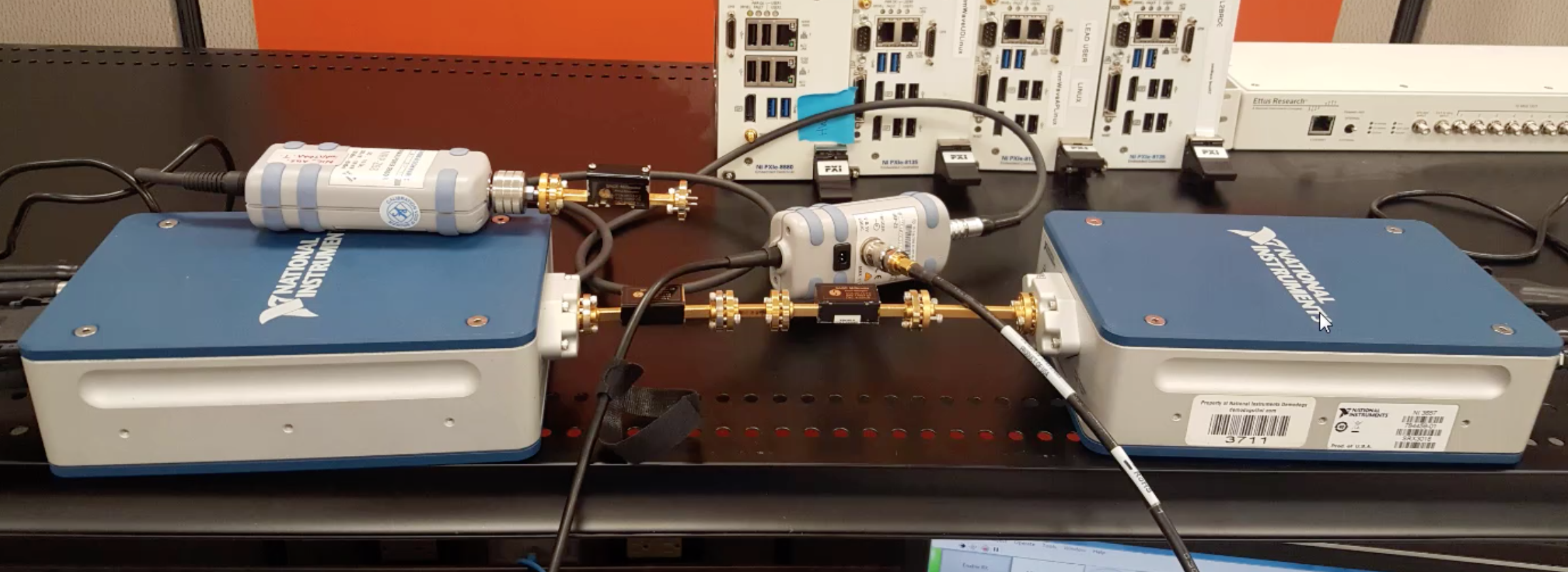}}
%  \vspace{1.5cm}
  \centerline{(a) Calibration set-up}\medskip
\end{minipage}
%\hspace{0.05cm}
%\hfill
\begin{minipage}[b]{\linewidth}
  \centering
  \centerline{\includegraphics[clip,width=0.65\columnwidth]{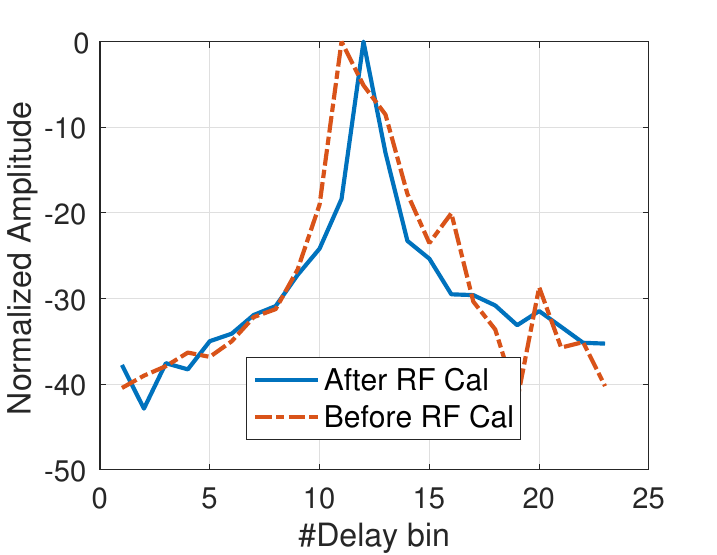}}
%  \vspace{1.5cm}
  \centerline{(b) Corner reflector response}\medskip
\end{minipage}
\vspace{-2em}
\caption{RF hardware frequency response calibration set-up (left) and corner reflector response after and before RF hardware calibration. The delay bin resolution is the symbol period. To collect RF hardware frequency response for calibration, we used two STA-30-12-F2 attenuators of 30 dB each between the transmitter and receiver to avoid RX damage.}
\label{fig:RFCal}
%\vspace{-1.2em}
\end{figure}

 To be able to estimate the radar or communication channel accurately, we first measure the RF hardware channel response using loop-back mode and then we need to remove the impact of the hardware, which is known as calibration. In the loop-back mode, we connect the mmWave head of the transmitter and the receiver back-to-back with a 60 dB attenuator in between to measure the hardware frequency response as shown in Fig.~\ref{fig:RFCal}(a). We assume that the major non-flatness in the frequency response is caused by the TX-RX RF chain up to the mmWave head. This setup, however, assumes negligible non-flatness due to the RF frequency response of the waveguide connectors and passive components, such as horn antennas, themselves have flat response. Then, we equalize the hardware frequency response using frequency-domain MMSE equalizer.

Fig.~\ref{fig:RFCal}(b) shows the estimated channel response of a trihedral corner reflector of 4.3 inch mounted on a toy car before the equalization. This channel response deviates from the ideal narrow thumbtack shape. Therefore, we applied the RF hardware equalization and also changed the base of the corner reflector with narrow wooden flat surface to achieve near ideal thumbtack shape of the channel response. We see that equalized channel response has narrower mainlobe width and smaller sidelobes as compared to the unequalized one.

  %%%%%%%%%%%%%%
\subsubsection{Estimated radar channels in the range-azimuth domain}

     \begin{figure}[!h]
\centering
\includegraphics[width=0.65\columnwidth]{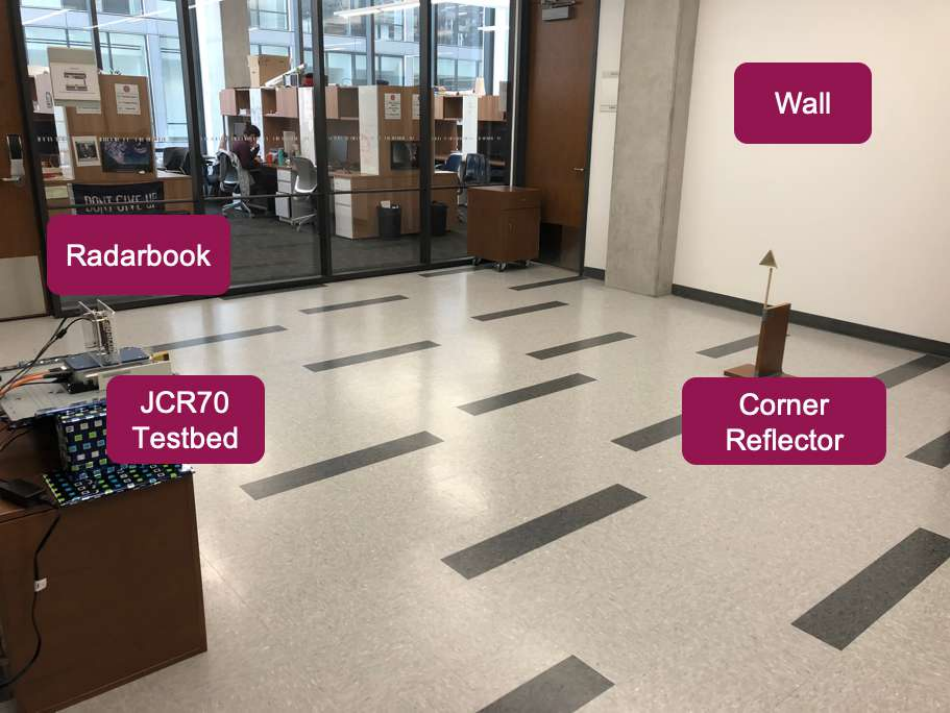}
 \caption{Experimental set-up to evaluate the radar performance of our testbed for a single-target scenario using a corner reflector in the indoor lab.}
\label{fig:SceneSingle}
\vspace{-1.2em}
\end{figure}

  \begin{figure}[tbp]
\begin{minipage}[b]{\linewidth}
  \centering
  \centerline{ \includegraphics[clip,width=0.65\columnwidth]{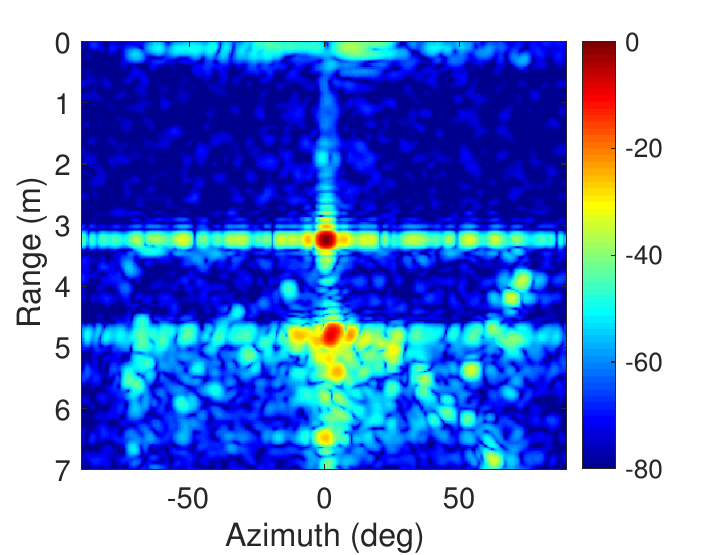}}
%  \vspace{1.5cm}
  \centerline{(a) JCR70 traditional processing}\medskip
\end{minipage}
%\hspace{0.05cm}
%\hfill
\begin{minipage}[b]{\linewidth}
  \centering
  \centerline{\includegraphics[clip,width=0.65\columnwidth]{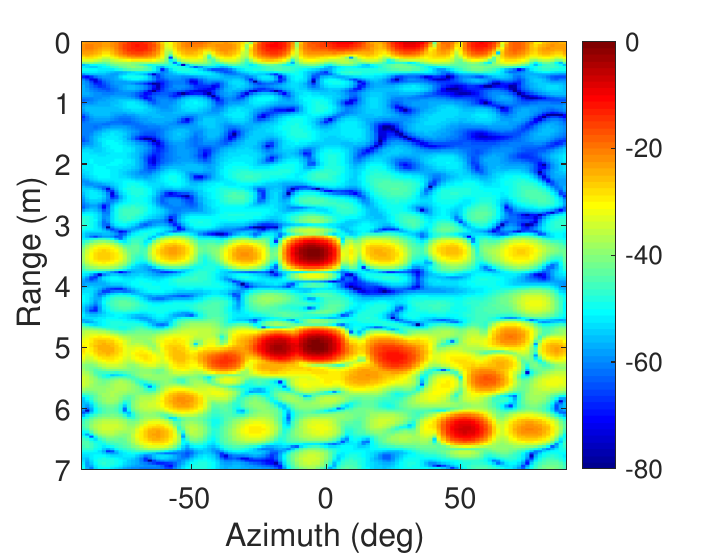}}
%  \vspace{1.5cm}
  \centerline{(b) Radarbook traditional processing}\medskip
\end{minipage}
\begin{minipage}[b]{0.49\linewidth}
  \centering
  \centerline{ \includegraphics[clip,width=\columnwidth]{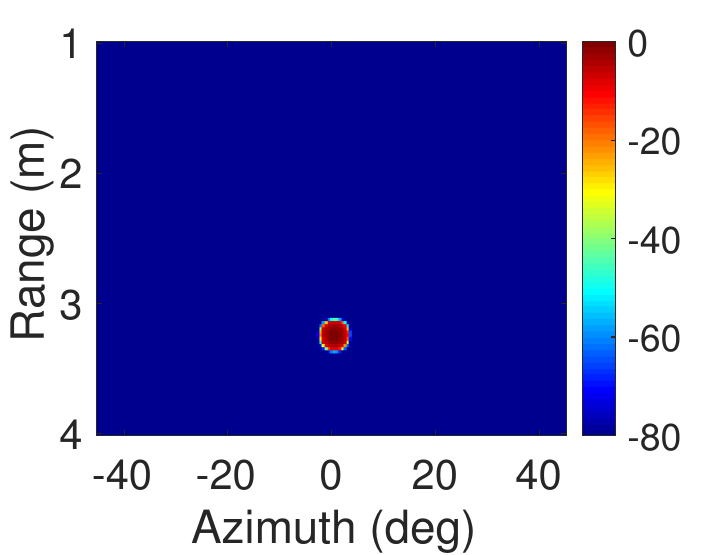}}
%  \vspace{1.5cm}
  \centerline{(c) JCR70 GAMP}\medskip
\end{minipage}
%\hspace{0.05cm}
%\hfill
\begin{minipage}[b]{0.49\linewidth}
  \centering
  \centerline{\includegraphics[clip,width=\columnwidth]{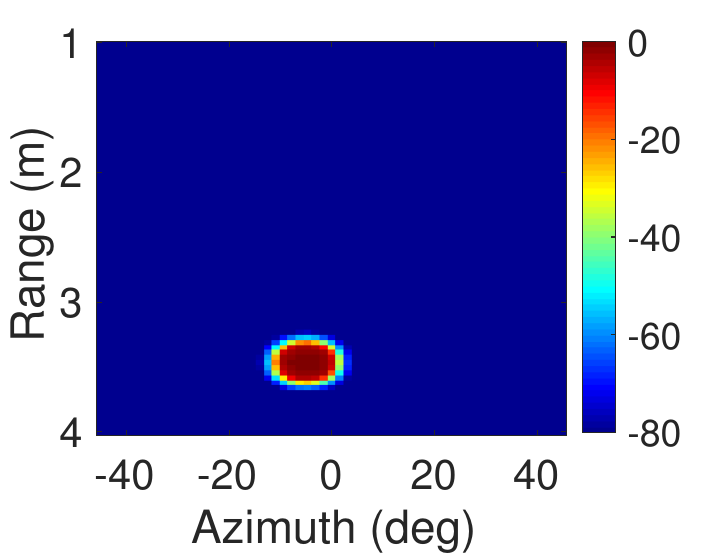}}
%  \vspace{1.5cm}
  \centerline{(d) Radarbook GAMP}\medskip
\end{minipage}
%\vspace{-1.2cm}
\caption{Estimated radar channels for a single-target scenario with traditional and advanced processing using our testbed (left) as well as the Radarbook (right). The channel estimates in (c/d) with EM-BG-GAMP have reduced sidelobes than (a/b) with traditional processing.}
\label{fig:FFT_GAMP_Single}
\vspace{-1.6em}
\end{figure}

\begin{figure}[htb]
\begin{minipage}[b]{\linewidth}
  \centering
  \centerline{ \includegraphics[clip,width=0.65\columnwidth]{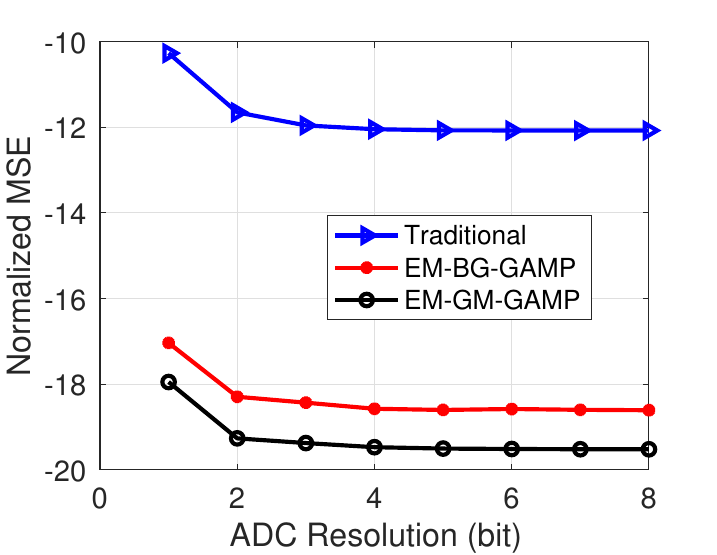}}
%  \vspace{1.5cm}
  \centerline{(a) NMSE versus ADC resolution}\medskip
\end{minipage}
%\hspace{0.05cm}
%\hfill
\begin{minipage}[b]{\linewidth}
  \centering
  \centerline{\includegraphics[clip,width=0.65\columnwidth]{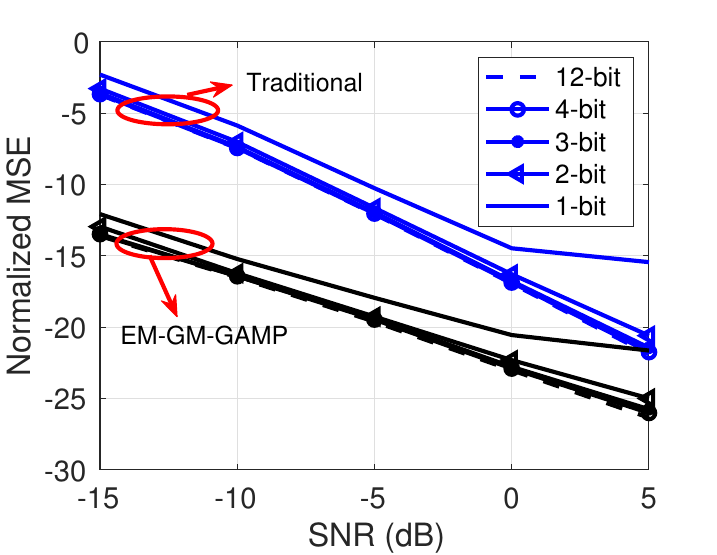}}
%  \vspace{1.5cm}
  \centerline{(b) NMSE versus SNR}\medskip
\end{minipage}
 \vspace{-0.7cm}
\caption{Estimated radar NMSE for the single-target scenario versus ADC resolutions at SNR equal to -5 dB (left) and versus SNRs for 12-, 4-, 3-, 2-, as well as 1- bit  ADC resolutions (right). The radar channel estimate with 2-bit ADCs perform closely to the high-resolution ADCs at low and medium SNRs, whereas with 3-bit ADCs perform closely at all considered SNR values.}
\label{fig:Single_NADC}
\vspace{-1.2em}
\end{figure}

To evaluate the radar performance of our JCR testbed in the range-angle domain, we placed a corner reflector of 0.1~m edge length at 3.21~m and 0$^{\circ}$ with respect to our testbed, used a horn antenna with 10 dBi gain, and moved the TX with $M = 86$ steps. Additionally, we used the Radarbook to estimate the radar channel for comparison and placed it above our setup, in the indoor lab as shown in Fig.~\ref{fig:SceneSingle}.  For benchmarking purposes, we measured the target distance and angle using a laser device with mm-level accuracy. We also determined the SIMO antenna pattern of our developed JCR70 testbed using this set-up and compared it with the ideal antenna pattern. The resulting antenna pattern using our testbed was found to be close to the ideal pattern, and the detailed description on the JCR70 antenna pattern can be found in our paper~\cite{KumMazHea:A-MIMO-Joint-Communication-Radar:20}.

Figs.~\ref{fig:FFT_GAMP_Single}(a) and (b) shows the normalized radar channels estimated with traditional processing algorithm using our testbed and the Radarbook, while Figs.~\ref{fig:FFT_GAMP_Single}(c) and (d) show the estimated radar channels using EM-BG-GAMP algorithm with reduced sidelobes and noise. We observe the full-duplex effect in the traditional radar images. We used the tap corresponding to the self-interference effect in the channel estimate obtained at the first antenna location as the zero range reference. We found the tap corresponding to the corner reflector to be at a constant distance with respect to the zero reference for all $M$ steps, while its magnitude varied within 1~dB. 

From Fig.~\ref{fig:FFT_GAMP_Single}(a), we see that the corner reflector is at 3.223 m and 1$^{\circ}$, demonstrating high-resolution sensing capability of our testbed. We also observe wall reflection near 4.9 m and several multipath reflections around it. The wall reflection is stronger than the corner reflector because it has a larger radar cross-section and because the corner reflector was placed quite far from the boresight resulting in smaller antenna gain. From Figs.~\ref{fig:FFT_GAMP_Single}(c) and (d), we observe that the resolution of the single-target image using our testbed is much higher than the Radarbook due to higher bandwidth and the larger number of synthesized antennas.

    \begin{figure*}[!]
\begin{minipage}[b]{0.49\linewidth}
  \centering
  \centerline{ \includegraphics[clip,width=0.65\columnwidth]{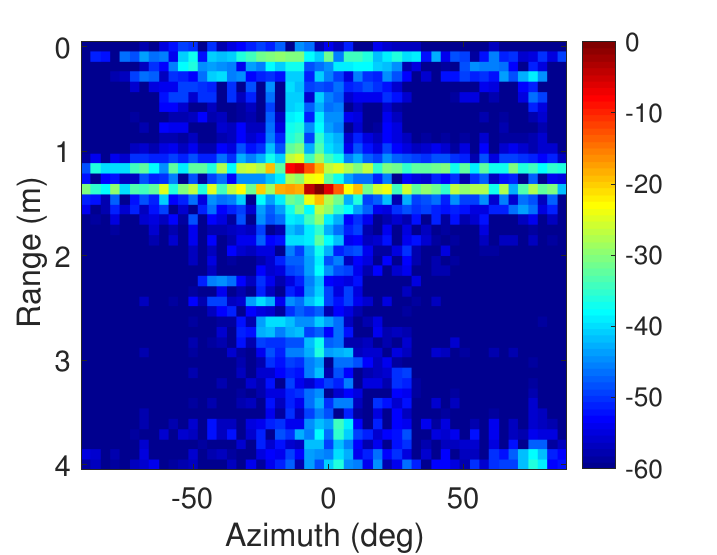}}
%  \vspace{1.5cm}
  \centerline{(a) JCR70 traditional processing}\medskip
\end{minipage}
%\hspace{0.05cm}
%\hfill
\begin{minipage}[b]{0.49\linewidth}
  \centering
  \centerline{\includegraphics[clip,width=0.65\columnwidth]{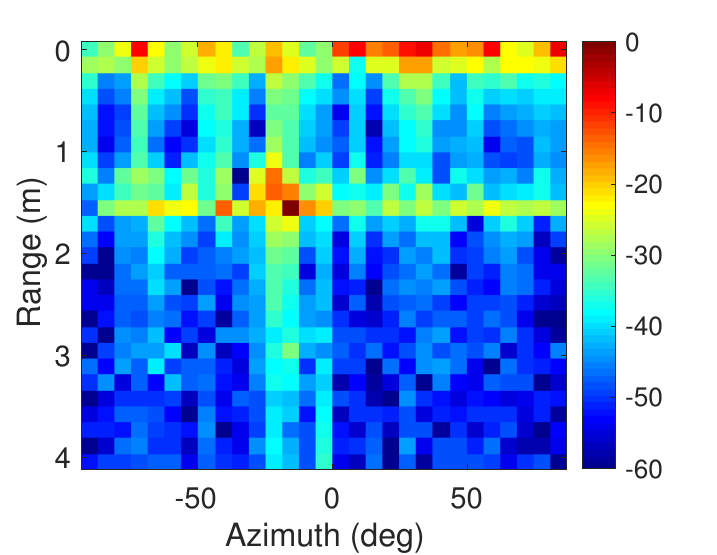}}
%  \vspace{1.5cm}
  \centerline{(b) Radarbook traditional processing}\medskip
\end{minipage}
\begin{minipage}[b]{0.49\linewidth}
  \centering
  \centerline{\includegraphics[clip,width=0.65\columnwidth]{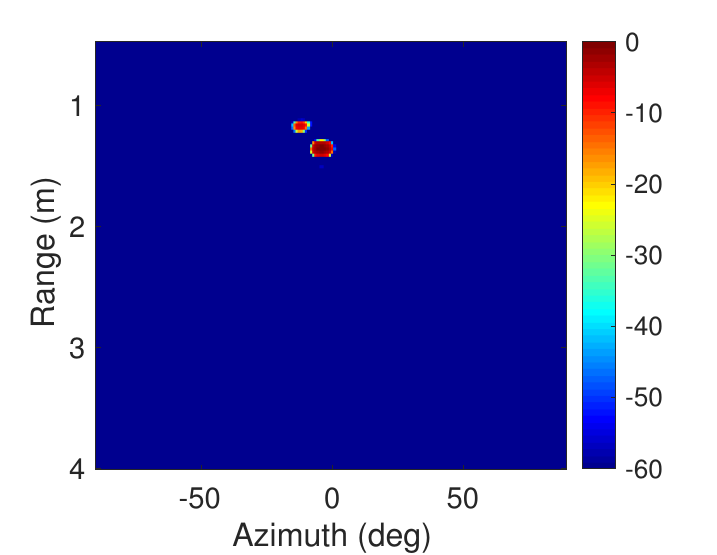}}
%  \vspace{1.5cm}
  \centerline{(c) JCR70 GAMP processing}\medskip
\end{minipage}
\begin{minipage}[b]{0.49\linewidth}
  \centering
  \centerline{\includegraphics[clip,width=0.65\columnwidth]{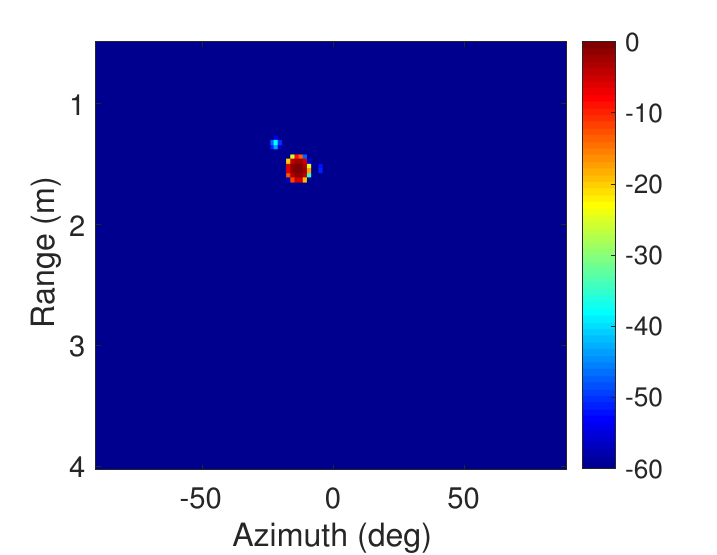}}
%  \vspace{1.5cm}
  \centerline{(d) Radarbook GAMP processing}\medskip
\end{minipage}
\vspace{-0.2cm}
\caption{Estimated radar channels for the two-target scenario using corner reflectors of 4.3 inch and 3.2 inch edge length with traditional and advanced processing using our testbed (left) as well as the Radarbook (right). The channel estimates in (c/d) with EM-BG-GAMP have reduced sidelobes than (a/b) with traditional processing. The channel estimates in (a/c) using our JCR70 testbed have recovered the two-target channel response better than the channel estimates in (b/d) using the Radarbook.}
\label{fig:FFT_GAMP_Two}
\vspace{-1.1em}
\end{figure*}

Figs.~\ref{fig:Single_NADC}(a) and (b) show the estimated radar NMSE variation with different ADC resolutions and SNRs for the single-target experiment using emulations on the data collected from our testbed. Fig.~\ref{fig:Single_NADC}(a) shows the estimated NMSE variation with different ADC resolutions  at -5 dB, using the traditional algorithm, EM-BG-GAMP, and EM-GM-GAMP.  The estimated NMSE decreases marginally till 2 bit ADC. The gap between the estimated NMSEs of any two consecutive ADC resolution is highest between the 1-bit ADC and 2-bit ADC. Fig.~\ref{fig:Single_NADC}(b) depicts the estimated NMSE variation with different SNRs for 12-bit, 4-bit, 3-bit, 2-bit, and 1-bit ADCs using the traditional FFT-based algorithm and EM-GM-GAMP technique. The gap between high-resolution ADCs and low-resolution ADCs increases with SNR. The gap between the estimated NMSEs of traditional processing and sparse GAMP technique decreases with SNR. From Figs.~\ref{fig:Single_NADC}(a) and (b), we see that 2-bit ADCs perform very closely to the high-resolution ADCs at low and medium SNRs, whereas 3-bit ADCs perform very closely at all considered SNR values.

%%%%%%%%%%%%%%%%%%%%%%%%%%%%
\subsection{Radar: Two-target scenarios}
%%%%%%%%%%%%%%%%%%%%%%%%%%%% 
  \begin{figure}[tbp]
\centering
\includegraphics[width=0.4\columnwidth]{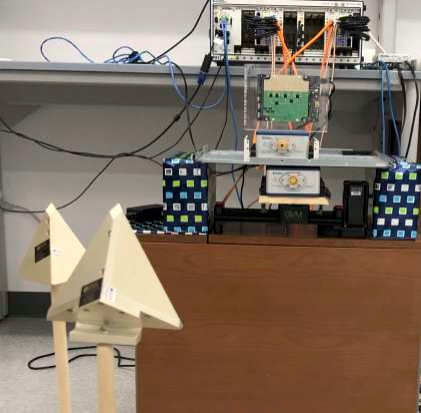}
 \caption{Experimental set-up to evaluate the radar performance of our testbed for a two-target scenario using two corner reflectors in the indoor lab.}
\label{fig:SceneTwo}
\vspace{-1.2em}
\end{figure}

\begin{figure}[tbp]
\begin{minipage}[b]{\linewidth}
  \centering
  \centerline{ \includegraphics[clip,width=0.65\columnwidth]{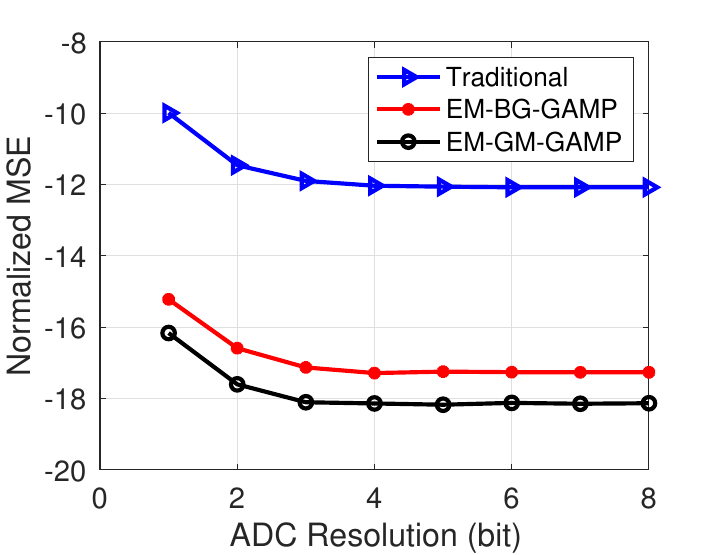}}
%  \vspace{1.5cm}
  \centerline{(a) NMSE versus ADC resolution}\medskip
\end{minipage}
%\hspace{0.05cm}
%\hfill
\begin{minipage}[b]{\linewidth}
  \centering
  \centerline{\includegraphics[clip,width=0.65\columnwidth]{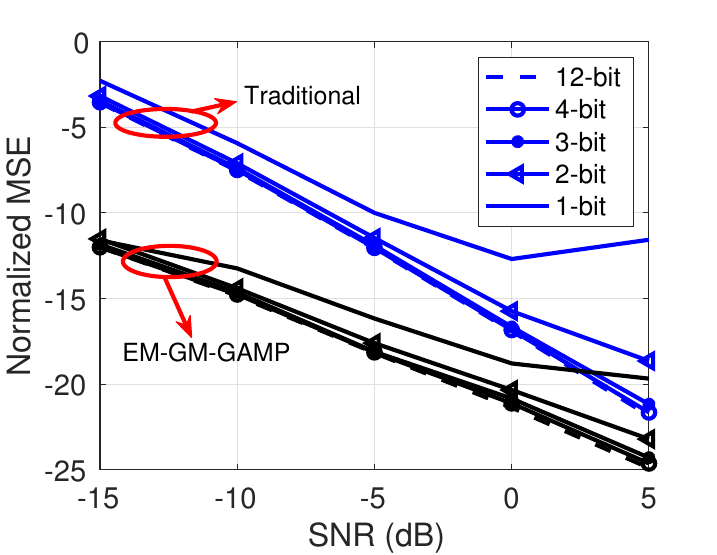}}
%  \vspace{1.5cm}
  \centerline{(b) NMSE versus SNR}\medskip
\end{minipage}
\vspace{-0.7cm}
\caption{Estimated radar NMSE for the two-target scenario for different ADC resolutions at SNR equal to -5 dB (left) and versus SNRs for 12-, 4-, 3-, 2-, as well as 1- bit  ADC resolutions (right).  The radar channel estimate with 2-bit ADCs perform closely to the high-resolution ADCs at low SNRs, whereas with 3-bit ADCs perform closely at all considered SNR values.}
\label{fig:Two_NADC}
%\vspace{-1.2cm}
\end{figure}

  \begin{figure*}[tbp]
\begin{minipage}[b]{0.49\linewidth}
  \centering
  \centerline{ \includegraphics[clip,width=0.65\columnwidth]{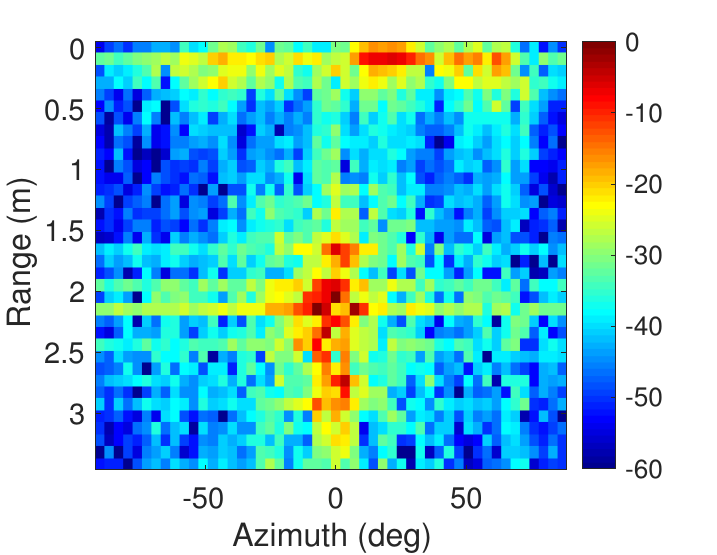}}
%  \vspace{1.5cm}
  \centerline{(a) JCR70 traditional processing}\medskip
\end{minipage}
%\hspace{0.05cm}
%\hfill
\begin{minipage}[b]{0.49\linewidth}
  \centering
  \centerline{\includegraphics[clip,width=0.65\columnwidth]{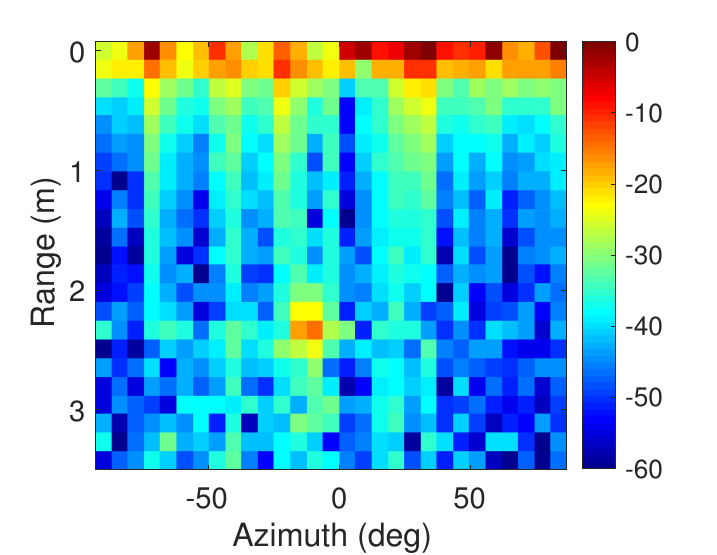}}
%  \vspace{1.5cm}
  \centerline{(b) Radarbook traditional processing}\medskip
\end{minipage}
\hspace{0.05cm}\begin{minipage}[b]{0.49\linewidth}
  \centering
  \centerline{ \includegraphics[clip,width=0.65\columnwidth]{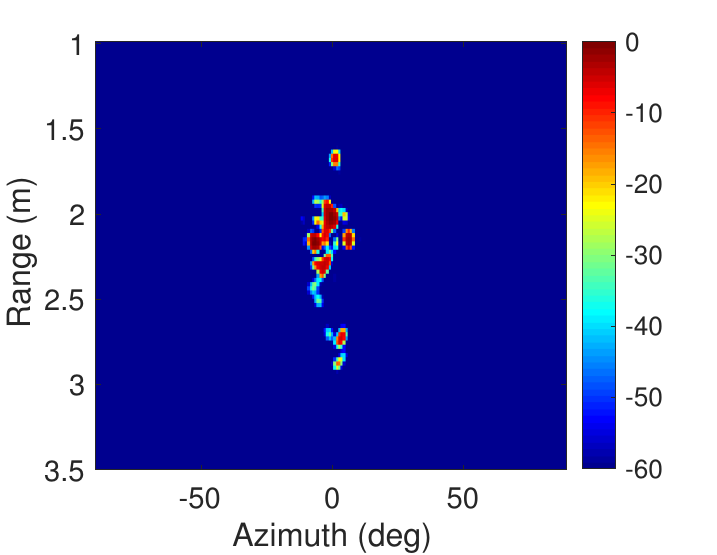}}
%  \vspace{1.5cm}
  \centerline{(c) JCR70 GAMP processing}\medskip
\end{minipage}
%\hspace{0.05cm}
%\hfill
\begin{minipage}[b]{0.49\linewidth}
  \centering
  \centerline{\includegraphics[clip,width=0.65\columnwidth]{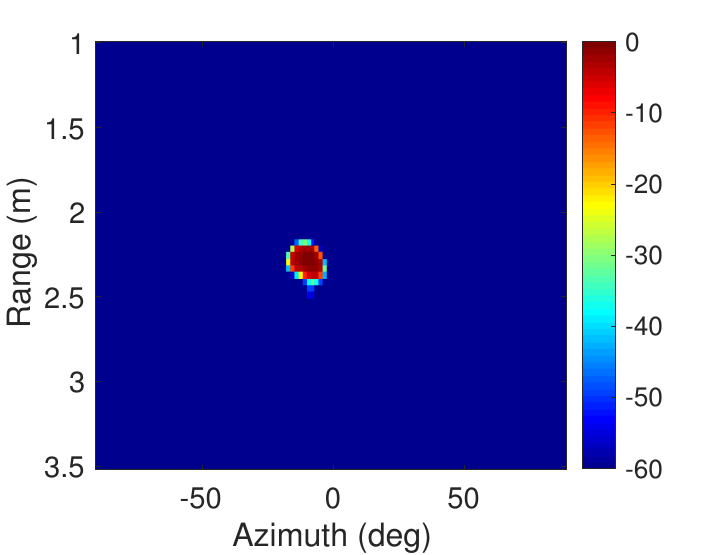}}
%  \vspace{1.5cm}
  \centerline{(d) Radarbook GAMP processing}\medskip
\end{minipage}
\vspace{-0.2cm}
\caption{Estimated radar channels for the indoor bike scenario with traditional and advanced processing using our testbed (left) as well as the Radarbook (right). The channel estimates in (c/d) with EM-BG-GAMP have reduced sidelobes than (a/b) with traditional processing. The channel estimates in (a/c) using our JCR70 testbed have resolved the bike much better than the channel estimates in (b/d) using the Radarbook.}
\label{fig:ExtBike}
\vspace{-1.0em}
\end{figure*}

We performed a two-target experiment with two corner reflectors of 4.3 inch and 3.2 inch edge length in the indoor lab using our fully-digital SIMO testbed and the Radarbook. The two targets are closely placed in range and angle domain. Figs.~\ref{fig:FFT_GAMP_Two}(a) and (b) shows the estimated radar channels with traditional processing algorithm using our testbed and the Radarbook, while  Figs.~\ref{fig:FFT_GAMP_Two}(c) and (d) show the estimated radar channels using EM-BG-GAMP algorithm with reduced sidelobes and noise. Fig.~\ref{fig:SceneTwo} shows the experimental set-up for the two-target scenario. We use 30 steps to emulate SIMO using our testbed. In Fig.~\ref{fig:FFT_GAMP_Two}(a), we observe two scattering centers corresponding to the two corner reflectors used, unlike the Radarbook due to higher bandwidth and number of synthesized antennas. The GAMP processed image in (c) has recovered the amplitudes of the two corner reflectors better than the one in (d).

Figs.~\ref{fig:Two_NADC}(a) and (b) show the estimated radar NMSE variation with different ADC resolutions and SNRs for the two-target experiment using emulations on the data collected from our testbed. Fig.~\ref{fig:Two_NADC}(a) shows the estimated NMSE variation with different ADC resolutions at -5 dB, using the traditional algorithm, EM-BG-GAMP, and EM-GM-GAMP. The estimated NMSE decreases marginally till 3 bit ADC. The gap between the estimated NMSEs of any two consecutive ADC resolution is highest between the 1-bit ADC and 2-bit ADC. The gap between traditional FFT-based processing and advanced GAMP algorithms is smaller as compared to single-target scenarios. Fig.~\ref{fig:Two_NADC}(b) depicts the estimated NMSE variation with different SNRs for 12-bit, 4-bit, 3-bit, 2-bit, and 1-bit ADCs using traditional FFT-based algorithm and EM-GM-GAMP technique. The gap between high-resolution ADCs and low-resolution ADCs increases with SNR. The gap between the estimated NMSEs of traditional processing and sparse GAMP technique decreases with SNR for all considered ADC resolutions except 1-bit ADC, where it first decreases and then increases with SNR. From Figs.~\ref{fig:Two_NADC}(a) and (b), we see that 2-bit ADCs perform very closely to the high-resolution ADCs at low SNRs, and 3-bit ADCs perform very closely at all considered SNR values.  
 %%%%%%%%%%%%%
\subsection{Radar: Extended target scenarios}
  %%%%%%%%%%%%%

\begin{figure}[tbp]
\centering
\includegraphics[width=0.65\columnwidth]{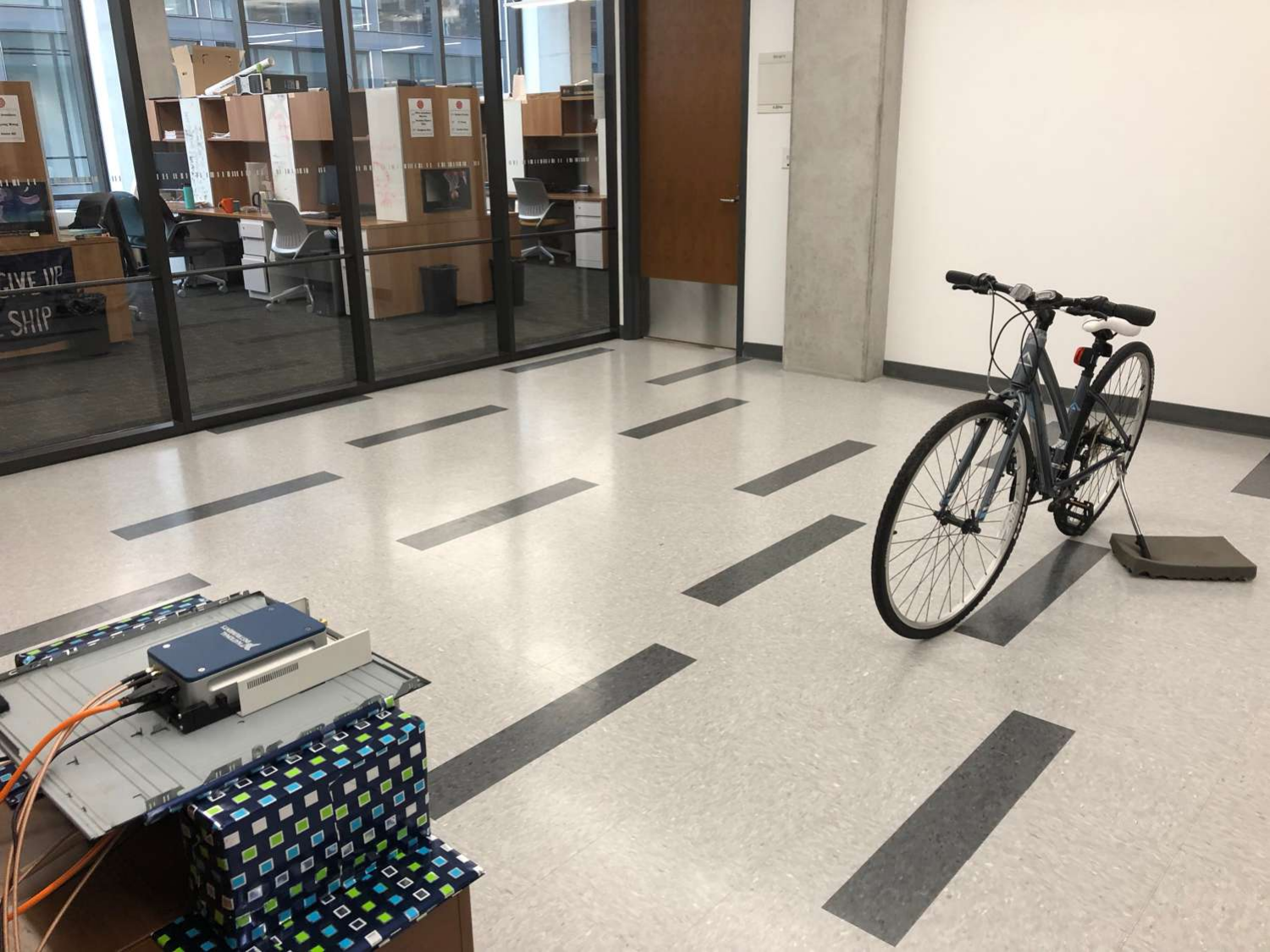}
 \caption{Experimental set-up to evaluate the radar performance of our testbed using a bike in the indoor lab.}
\label{fig:SceneBike}
\vspace{-0.3cm}
\end{figure}

We used a bike in the indoor lab to evaluate the radar performance in the extended target scenario using our fully-digital SIMO testbed and the Radarbook. Figs.~\ref{fig:ExtBike}(a) and (b) shows the estimated radar channels with traditional processing algorithm using our testbed and the Radarbook, while  Figs.~\ref{fig:ExtBike}(c) and (d) show the estimated radar channels using EM-BG-GAMP algorithm with reduced sidelobes and noise. Fig.~\ref{fig:SceneBike} shows the experimental set-up for the extended target scenario using a bike.  We use 50 steps to emulate SIMO using our testbed. In Fig.~\ref{fig:ExtBike}(c), we observe that multiple scattering centers corresponding to different parts of the bike. Due to higher bandwidth and number of synthesized antennas, the resolution of the bike image using our testbed is much higher than the INRAS Radarbook.

\begin{figure}[tbp]
\vspace{-0.2cm}
\begin{minipage}[b]{\linewidth}
  \centering
  \centerline{ \includegraphics[clip,width=0.65\columnwidth]{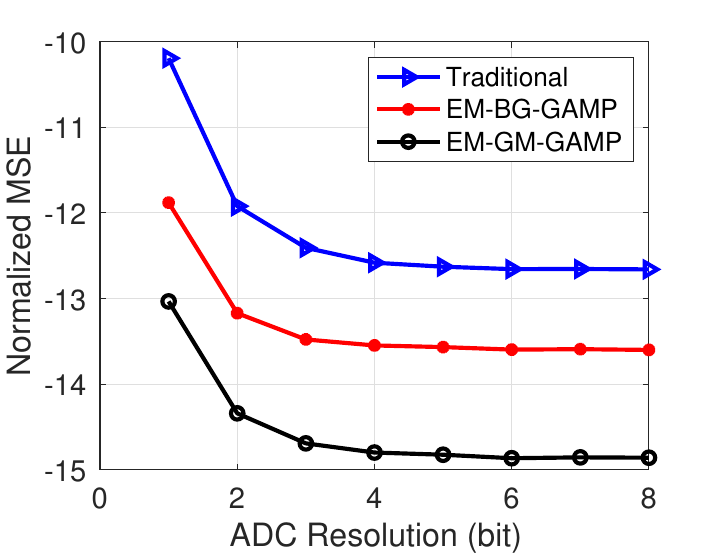}}
%  \vspace{1.5cm}
  \centerline{(a) NMSE versus ADC resolution}\medskip
\end{minipage}
\hspace{0.05cm}
%\hfill
\begin{minipage}[b]{\linewidth}
  \centering
  \centerline{\includegraphics[clip,width=0.65\columnwidth]{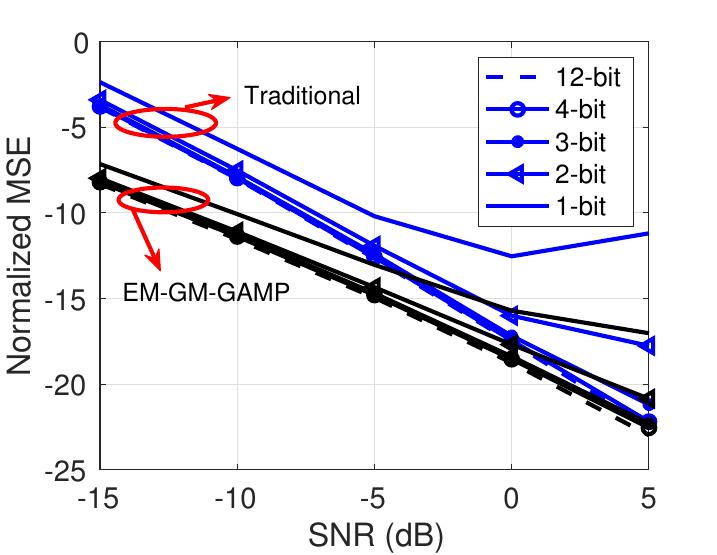}}
%  \vspace{1.5cm}
  \centerline{(b) NMSE versus SNR}\medskip
\end{minipage}
 \vspace{-0.7cm}
\caption{Estimated radar NMSE for the bike experiment for different ADC resolutions at SNR equal to -5 dB (left) and versus SNRs for 12-, 4-, 3-, 2-, as well as 1- bit  ADC resolutions (right). The radar channel estimate with 2-bit ADCs perform closely to the high-resolution ADCs at low SNR, whereas with 3-bit ADCs perform closely for the medium SNR, and with 4-bit ADCs perform closely at all considered SNR values.}
\label{fig:Ext_Bike_NADC}
\vspace{-1.0em}
\end{figure}

Figs.~\ref{fig:Ext_Bike_NADC}(a) and (b) show the estimated radar NMSE variation with different ADC resolutions and SNRs for the bike experiment using emulations on the data collected from our testbed. Fig.~\ref{fig:Ext_Bike_NADC}(a) shows the estimated NMSE variation with different ADC resolutions at -5 dB, using the traditional algorithm, EM-BG-GAMP, and EM-GM-GAMP.  The estimated NMSE decreases marginally till 4 bit ADC. The gap between the estimated NMSEs of any two consecutive ADC resolution is highest between the 1-bit ADC and 2-bit ADC. The gap between traditional processing and advanced GAMP algorithms is smaller as compared to single- and two-target scenarios. Fig.~\ref{fig:Ext_Bike_NADC}(b) depicts the estimated NMSE variation with different SNRs for 12-bit, 4-bit, 3-bit, 2-bit, and 1-bit ADCs using traditional FFT-based algorithm and EM-GM-GAMP technique. The gap between high-resolution ADCs and low-resolution ADCs increases with SNR. The gap between the estimated NMSEs of traditional processing and sparse GAMP technique for high-resolution ADCs decreases with SNR, whereas it first decreases and then increases with SNR for the low-resolution ADCs. From Figs.~\ref{fig:Ext_Bike_NADC}(a) and (b), we see that 2-bit ADCs perform very closely to the high-resolution ADCs at low SNR, 3-bit ADCs perform very closely for the medium SNR, and 4-bit ADCs perform very closely at all considered SNR values.

  \begin{figure}[tbp]
\begin{minipage}[b]{\linewidth}
  \centering
  \centerline{ \includegraphics[clip,width=0.65\columnwidth]{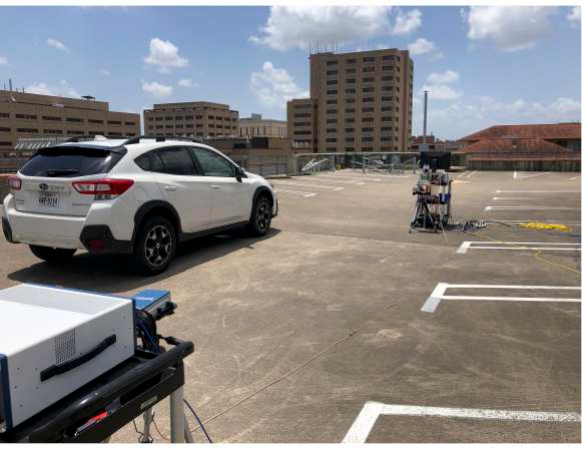}}
%  \vspace{1.5cm}
 % \centerline{(a) Outdoor scenario}\medskip
\end{minipage}
%\hspace{0.05cm}
%\hfill
%\begin{minipage}[b]{\linewidth}
%  \centering
%  \centerline{\includegraphics[clip,width=0.38\columnwidth]{Fig/Figure18b}}
%%  \vspace{1.5cm}
%  \centerline{(b) Radar set-up}\medskip
%\end{minipage}
 \vspace{-0.7cm}
\caption{Outdoor JCR measurement scenario using a car on the top of a parking garage. The cart with the JCR70 transmitter and radar receiver is kept in front of the car, whereas the cart with the JCR70 communication receiver is kept behind the car. }
\label{fig:SceneOutdoorJCR}
%\vspace{-1.0em}
\end{figure}

 \begin{figure}[tbp]
\begin{minipage}[b]{\linewidth}
  \centering
  \centerline{ \includegraphics[clip,width=0.6\columnwidth]{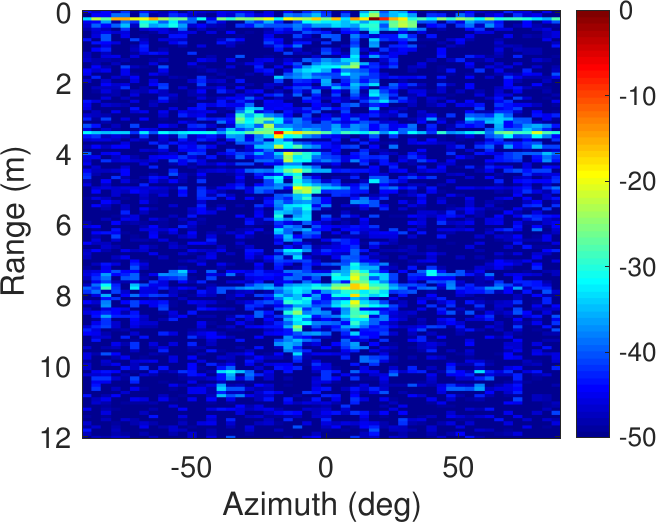}}
%  \vspace{1.5cm}
  \centerline{(a) Traditional processing}\medskip
\end{minipage}
%\hspace{0.05cm}
%\hfill
\begin{minipage}[b]{\linewidth}
  \centering
  \centerline{\includegraphics[clip,width=0.6\columnwidth]{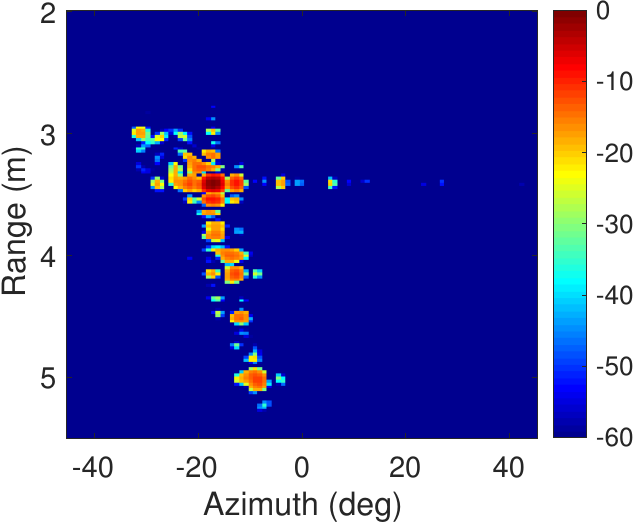}}
%  \vspace{1.5cm}
  \centerline{(b) GAMP processing}\medskip
\end{minipage}
 \vspace{-0.7cm}
\caption{Estimated radar channels for the outdoor car scenario with traditional and advanced processing using our testbed (left) and the Radarbook (right). The channel estimates in (a) have reduced sidelobes than (b).  Due to the high-resolution of our JCR70 testbed, the channel estimate in (b) shows several scattering centers corresponding to the car.}
\label{fig:FFT_GAMP_Car}
\vspace{-1.6em}
\end{figure}

Additionally, we also conduct outdoor joint communication and radar experiments with 50 steps using Subaru Crosstrek as the vehicle target on Speedway parking garage in UT Austin, as shown in Fig.~\ref{fig:SceneOutdoorJCR}. From Fig.~\ref{fig:SceneOutdoorJCR} and Fig.~\ref{fig:FFT_GAMP_Car}(b), we see that the JCR transmitter and the communication receiver were separated by 7.51~m. The vehicle target was placed in between the JCR transmitter and the communication receiver. The directivity of the horn antenna reduced the reflections from the railings and poles around our set-up. Fig.~\ref{fig:FFT_GAMP_Car}(a) shows the estimated radar channel around the car reflections using the traditional FFT-based algorithm with an interpolation of factor 4, while Fig.~\ref{fig:FFT_GAMP_Car}(b) shows the estimated radar channel using the advanced EM-BG-GAMP technique. We see that Fig.~\ref{fig:FFT_GAMP_Car}(b) has reduced sidelobes and noise as compared to Fig.~\ref{fig:FFT_GAMP_Car}(a). The range-spread of the car is much wider in range than that of the bike in Fig.~\ref{fig:ExtBike}. Due to the high-resolution of our testbed, we can also see multiple scattering centers corresponding to the car which makes the car radar image look quite different than the bike.

%%%%%%%%%%%%%%%%%%%%%%%%%%
\subsection{Joint communication-radar}
%%%%%%%%%%%%%%%%%%%%%%%%%%

\begin{figure}[tbp]
\begin{minipage}[b]{\linewidth}
  \centering
  \centerline{ \includegraphics[clip,width=0.65\columnwidth]{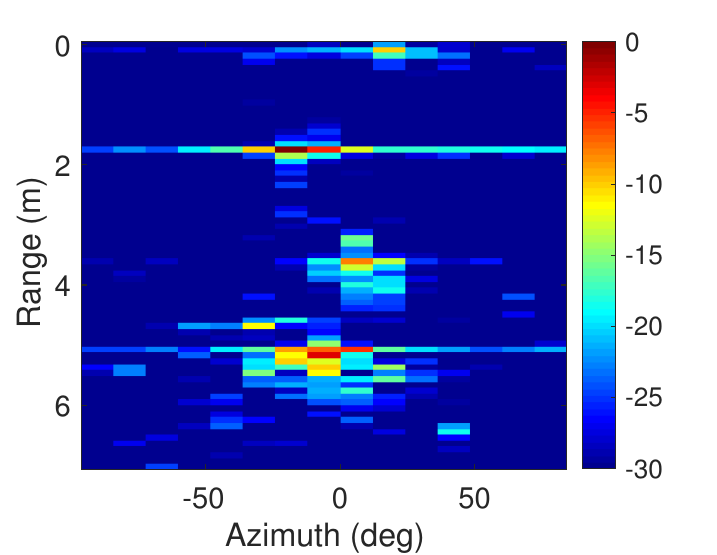}}
%  \vspace{1.5cm}
  \centerline{(a) Radar}\medskip
\end{minipage}
%\hspace{0.05cm}
%\hfill
\begin{minipage}[b]{\linewidth}
  \centering
  \centerline{\includegraphics[clip,width=0.65\columnwidth]{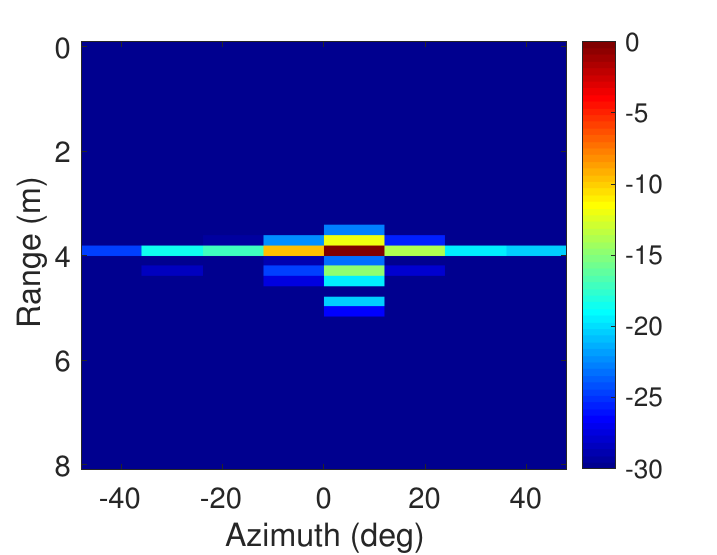}}
%  \vspace{1.5cm}
  \centerline{(b) Communication}\medskip
\end{minipage}
\vspace{-0.7cm}
\caption{Estimated radar channel (left) and communication channel (right) using our JCR70 testbed. In the radar channel estimate, the full-duplex effect is observed around 0~m along with reflections from surrounding objects, such as the communication receiver at 1.75~m. In the communication channel, the LoS path between the communication TX and RX is observed at 1.75 m.}
\label{fig:FFT_JCR}
\vspace{-0.7em}
\end{figure}

The performance of our fully-digital SIMO wideband testbed is also evaluated for the simultaneous communication and radar modes at 73 GHz. We conducted JCR experiments with 15 steps in the same indoor lab as in \figref{fig:SceneSingle}.

 \begin{figure}[tbb]
\begin{minipage}[b]{\linewidth}
  \centering
  \centerline{ \includegraphics[clip,width=0.65\columnwidth]{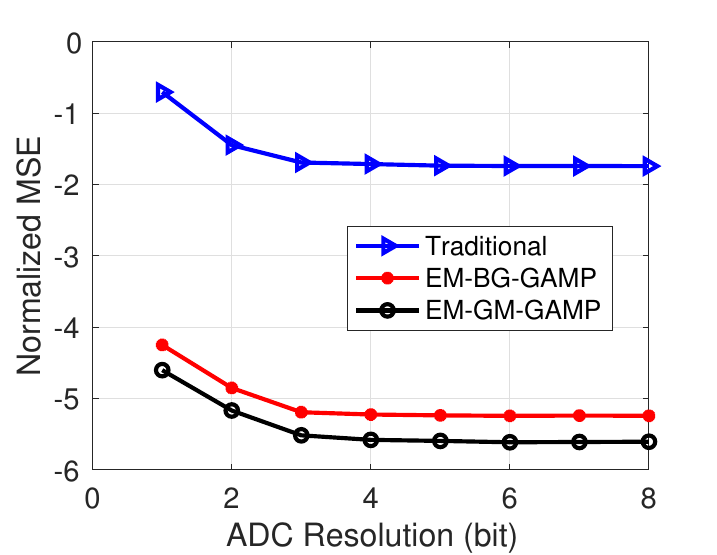}}
%  \vspace{1.5cm}
  \centerline{(a) Radar}\medskip
\end{minipage}
%\hspace{0.05cm}
%\hfill
\begin{minipage}[b]{\linewidth}
  \centering
  \centerline{\includegraphics[clip,width=0.65\columnwidth]{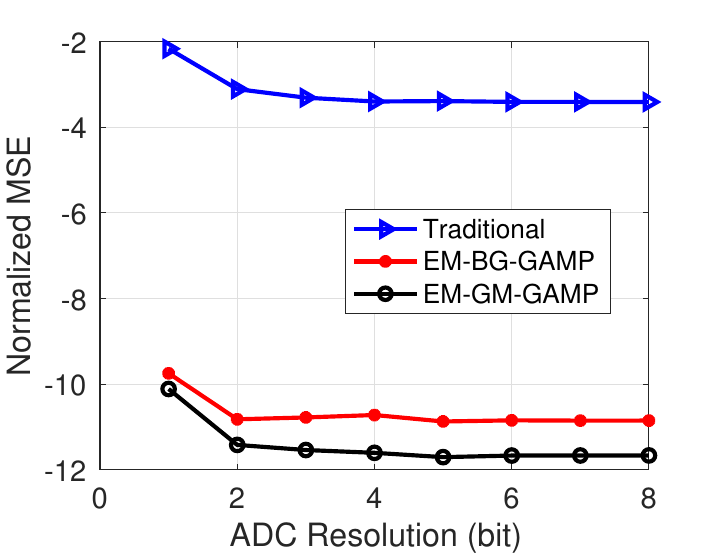}}
%  \vspace{1.5cm}
  \centerline{(b) Communication}\medskip
\end{minipage}
\vspace{-0.7cm}
\caption{Estimated NMSEs for radar (left) and communication (right) with different ADC resolutions at SNR of -15 dB. The JCR estimate with 3-bit ADCs perform closely to the high-resolution ADCs case.}
\label{fig:JCR_NADC_bit}
\vspace{-0.7em}
\end{figure}

Fig.~\ref{fig:FFT_JCR}(a) shows the estimated indoor radar channel in the range-azimuth domain using the traditional processing, while Fig.~\ref{fig:FFT_JCR}(b) shows the estimated communication channel in the outdoor setting. In the estimated radar channel image, we observe the full-duplex effect unlike the estimated communication channel image. The delay-spread in the radar channel is found to be higher than the communication channel because of the long-distance targets. In Fig.~\ref{fig:FFT_JCR}(a), we observe that the direct path corresponding to the communication receiver is at 1.75~m and -8.4 degrees. The wall reflection is more spread as compared to the communication receiver around 4.98~m because of its spatial extent and strong multipath effect due to the large radar cross-section. We also see the reflection of the metallic chassis at 3.613~m. The angular resolution and dynamic range of the JCR image in Fig.~\ref{fig:FFT_JCR} is worse than the earlier traditional FFT-processed radar images, such as in Fig.~\ref{fig:FFT_GAMP_Car}, due to the lower number of antenna steps used in this JCR experiment. A detailed analysis of this JCR setting with high-resolution ADCs and the BG-GAMP processing algorithm can be found in our conference paper~\cite{KumMazHea:A-MIMO-Joint-Communication-Radar:20}.

In Figs.~\ref{fig:JCR_NADC_bit}(a) and (b), we also compare the JCR performance using the NMSE metric for different ADC resolutions at an SNR of -15 dB for the indoor setting. Due to the lower radar SNR resulting from the two-way radar channel as compared to the one-way communication channel, we see that radar NMSE is worse than the communication NMSE. We also see that the gap between the traditional FFT-based method and the sparse GAMP method is lower in radar than the communication. This could be due to the smaller delay spread and sparser channel in the LoS communication channel as compared to the indoor radar channel. From Figs.~\ref{fig:JCR_NADC_bit}(a) and (b), we see that ADC resolution of 3 bit performs very closely to the high-resolution ADC case.

%%%%%%%%%%%%%%%%%%%%%%%%%%%%%%%%%%%%%%%
\section{Conclusion and discussion} \label{sec:Conclusion}
%%%%%%%%%%%%%%%%%%%%%%%%%%%%%%%%%%%%%%%

In this paper, we developed a low-complexity proof-of-concept platform for a fully-digital joint communication-radar sounding testbed with SIMO functionality and different ADC resolutions at 73 GHz carrier frequency with 2 GHz bandwidth. For the precise radar characterization of our JCR70 measurement platform, we conducted experiments for the single-target, two-target scenarios using trihedral corner reflectors. We also conducted experiments for extended target scenarios using a bike indoors and a car outdoors for demonstrating and evaluating the performance of our testbed. We applied both traditional FFT-based and advanced GAMP processing algorithms for JCR channel estimations. 

The results in this paper demonstrate the high-resolution capability with a wide field of view of our low-complexity wideband a fully-digital joint communication-radar testbed. The GAMP-based processing provided enhanced radar and communication channel estimates with reduced sidelobes and noise as compared to the traditional processing.  Our JCR70 platform with a fully-digital JCR waveform achieved higher resolution capability in the range-angle domain than the state-of-the-art automotive radar. A quantized receiver of 2-4 bit ADCs performed very closely to the high-resolution ADCs. The quantized receiver with 1-bit ADC performed closely to high-resolution ADCs at low SNR. The performance gap, however, grows with increasing SNR and decreasing channel sparsity. The normalized mean square errors for radar channel estimates were found higher than the communication channel estimates due to the two-way radar channel with colocated TX-RX instead of the one-way communication channel with widely separated TX-RX. 

The insights in this paper can be taken into account for designing a JCR waveform and developing receive processing for radar and communication with improved performance. The next step is to extend the mmWave MIMO JCR proof-of-concept development for a dynamic scenario, to demonstrate its capability for next-general high-performance automotive and drone applications.

\bibliographystyle{IEEEtran}
\bibliography{IEEEabrv,refProp1}

\begin{thebibliography}{10}
\providecommand{\url}[1]{#1}
\csname url@samestyle\endcsname
\providecommand{\newblock}{\relax}
\providecommand{\bibinfo}[2]{#2}
\providecommand{\BIBentrySTDinterwordspacing}{\spaceskip=0pt\relax}
\providecommand{\BIBentryALTinterwordstretchfactor}{4}
\providecommand{\BIBentryALTinterwordspacing}{\spaceskip=\fontdimen2\font plus
\BIBentryALTinterwordstretchfactor\fontdimen3\font minus
  \fontdimen4\font\relax}
\providecommand{\BIBforeignlanguage}[2]{{%
\expandafter\ifx\csname l@#1\endcsname\relax
\typeout{** WARNING: IEEEtran.bst: No hyphenation pattern has been}%
\typeout{** loaded for the language `#1'. Using the pattern for}%
\typeout{** the default language instead.}%
\else
\language=\csname l@#1\endcsname
\fi
#2}}
\providecommand{\BIBdecl}{\relax}
\BIBdecl

\bibitem{ChoVaGon:Millimeter-Wave-Vehicular-Communication:16}
J.~Choi, V.~Va, N.~Gonzalez-Prelcic, R.~Daniels, C.~R. Bhat, and R.~W. Heath,
  ``Millimeter-wave vehicular communication to support massive automotive
  sensing,'' \emph{{IEEE} Commun. Mag.}, vol.~54, no.~12, pp. 160--167, Dec.
  2016.

\bibitem{lien2016soli}
J.~Lien, N.~Gillian, M.~E. Karagozler, P.~Amihood, C.~Schwesig, E.~Olson,
  H.~Raja, and I.~Poupyrev, ``Soli: Ubiquitous gesture sensing with millimeter
  wave radar,'' \emph{ACM Trans. on Graph.}, vol.~35, no.~4, p. 142, Jul. 2016.

\bibitem{KumMazMez:Low-Resolution-Sampling-for-Joint:18}
P.~Kumari, K.~U. Mazher, A.~Mezghani, and R.~W. Heath, ``Low resolution
  sampling for joint millimeter-wave {MIMO} communication-radar,'' in
  \emph{Proc. IEEE Statistical Signal Process. Workshop}, Jun. 2018, pp.
  193--197.

\bibitem{AjoSreLoc:On-the-Feasibility-of-Using-IEEE:19}
H.~Ajorloo, C.~J. Sreenan, A.~Loch, and J.~Widmer, ``On the feasibility of
  using {IEEE 802.11ad} mmwave for accurate object detection,'' in \emph{Proc.
  ACM/SIGAPP Symp. on Appl. Comput.}, ser. SAC '19.\hskip 1em plus 0.5em minus
  0.4em\relax New York, NY, USA: ACM, Apr. 2019, pp. 2406--2413.

\bibitem{Kim:Experimental-Demonstration-of-MmWave:19}
W.~Kim, ``Experimental demonstration of mmwave vehicle-to-vehicle
  communications using {IEEE} 802.11ad,'' \emph{Sensors}, vol.~19, no.~9, May
  2019.

\bibitem{KumChoGon:IEEE-802.11ad-based-Radar::17}
P.~Kumari, J.~Choi, N.~Gonz{\'a}lez-Prelcic, and R.~W. Heath, ``{IEEE}
  802.11ad-based radar: An approach to joint vehicular communication-radar
  system,'' \emph{{IEEE} Trans. Veh. Technol.}, vol.~67, no.~4, pp. 3012--3027,
  Apr. 2018.

\bibitem{GroLopVen:Opportunistic-Radar-in-IEEE:18}
E.~Grossi, M.~Lops, L.~Venturino, and A.~Zappone, ``Opportunistic radar in
  {IEEE 802.11ad} networks,'' \emph{{IEEE} Trans. Signal Process.}, vol.~66,
  no.~9, pp. 2441--2454, May 2018.

\bibitem{RapR.WMur:Millimeter-Wave-Wireless:14}
T.~S. Rappaport, R.~W. Heath, J.~N. Murdock, and R.~C. Daniels,
  \emph{Millimeter Wave Wireless Communications}.\hskip 1em plus 0.5em minus
  0.4em\relax Pearson, 2014.

\bibitem{CudKovTho:Experimental-mm-wave-5G-cellular:14}
M.~{Cudak}, T.~{Kovarik}, T.~A. {Thomas}, A.~{Ghosh}, Y.~{Kishiyama}, and
  T.~{Nakamura}, ``Experimental mm wave {5G} cellular system,'' in \emph{Proc.
  IEEE Globecom Workshops}, Dec. 2014, pp. 377--381.

\bibitem{GomSisRib:Will-COTS-RF-Front-Ends:18}
R.~{Gomes}, L.~{Sismeiro}, C.~{Ribeiro}, T.~R. {Fernandes}, M.~G.
  {Sanch{\'e}z}, A.~{Hammoudeh}, and R.~F.~S. {Caldeirinha}, ``Will {COTS RF}
  front-ends really cope with {5G} requirements at {mmWave?}'' \emph{IEEE
  Access}, vol.~6, pp. 38\,745--38\,769, Jun. 2018.

\bibitem{LiuLuoXio:Low-Resolution-ADCs-for-Wireless:19}
J.~{Liu}, Z.~{Luo}, and X.~{Xiong}, ``Low-resolution {ADCs} for wireless
  communication: A comprehensive survey,'' \emph{IEEE Access}, vol.~7, pp.
  91\,291--91\,324, Jul. 2019.

\bibitem{MezAntNos:Multiple-parameter-estimation:10}
A.~Mezghani, F.~Antreich, and J.~A. Nossek, ``Multiple parameter estimation
  with quantized channel output,'' in \emph{Proc. Int. ITG Workshop on Smart
  Antennas}, Feb. 2010, pp. 143--150.

\bibitem{MoSchHea:Channel-Estimation-in-Broadband:18}
J.~{Mo}, P.~{Schniter}, and R.~W. {Heath}, ``Channel estimation in broadband
  millimeter wave {MIMO} systems with few-bit {ADCs},'' \emph{IEEE Trans. on
  Signal Process.}, vol.~66, no.~5, pp. 1141--1154, Mar. 2018.

\bibitem{ZahNagMod:One-Bit-Compressive-Radar:19}
S.~J. {Zahabi}, M.~M. {Naghsh}, M.~{Modarres-Hashemi}, and J.~{Li}, ``One-bit
  compressive radar sensing in the presence of clutter,'' \emph{IEEE Trans. on
  Aerosp. and Electronic Syst.}, pp. 1--1, May 2019.

\bibitem{MazMezHea:Low-Resolution-Millimeter-Wave:18}
K.~U. {Mazher}, A.~{Mezghani}, and R.~W. {Heath}, ``Low resolution millimeter
  wave radar: Bounds and performance,'' in \emph{Proc. Asilomar Conf. on
  Signals, Syst., and Comput.}, Oct. 2018, pp. 554--558.

\bibitem{niTestbed}
\BIBentryALTinterwordspacing
{NI mmWave} transceiver system. [Online]. Available:
  \url{http://www.ni.com/sdr/mmwave/}
\BIBentrySTDinterwordspacing

\bibitem{inras}
\BIBentryALTinterwordspacing
{INRAS product-Radarbook}. [Online]. Available:
  \url{http://www.inras.at/en/home.html}
\BIBentrySTDinterwordspacing

\bibitem{KumMazHea:A-MIMO-Joint-Communication-Radar:20}
P.~Kumari, A.~{Mezghani}, and R.~W. Heath, ``A {MIMO} joint communication-radar
  measurement platform at the millimeter-wave band,'' in \emph{Proc. Eur. Conf.
  in Antenna Propag.}, Mar. 2020.

\bibitem{KumMezHea:A-Low-Resolution-ADC-Proof-of-Concept-Development:20}
P.~{Kumari}, A.~{Mezghani}, and R.~W. {Heath}, ``A low-resolution {ADC}
  proof-of-concept development for a fully-digital millimeter-wave joint
  communication-radar,'' in \emph{Proc. IEEE Int. Conf. on Acoust., Speech and
  Signal Process.}, May 2020, pp. 8619--8623.

\bibitem{ieee2012wireless}
``{Wireless LAN Medium Access Control (MAC) and Physical Layer (PHY)
  Specifications. Amendment 3: Enhancements for Very High Throughput in the 60
  GHz Band},'' \emph{IEEE Std. 802.11ad}, 2012.

\bibitem{KabPat:Zadoff-Chu-Spreading-Sequence:20}
V.~B. Kaba and R.~R. Patil, ``{Zadoff Chu} spreading sequence for {5G} wireless
  communications,'' in \emph{Emerging Trends in Photonics, Signal Processing
  and Communication Engineering}.\hskip 1em plus 0.5em minus 0.4em\relax
  Springer, 2020, pp. 25--31.

\bibitem{RapMacSam:Wideband-Millimeter-Wave-Propagation:15}
T.~S. Rappaport, G.~R. MacCartney, M.~K. Samimi, and S.~Sun, ``Wideband
  millimeter-wave propagation measurements and channel models for future
  wireless communication system design,'' \emph{IEEE Trans. on Commun.},
  vol.~63, no.~9, pp. 3029--3056, Sep. 2015.

\bibitem{MacSunRap:Millimeter-Wave-Wireless:16}
G.~R. MacCartney, Jr., S.~Sun, T.~S. Rappaport, Y.~Xing, H.~Yan, J.~Koka,
  R.~Wang, and D.~Yu, ``Millimeter wave wireless communications: New results
  for rural connectivity,'' in \emph{Proc. Workshop on All Things Cellular:
  Operations, Applications and Challenges}, ser. ATC '16.\hskip 1em plus 0.5em
  minus 0.4em\relax New York, NY, USA: ACM, Oct. 2016, pp. 31--36.

\bibitem{bajwa2010compressed}
W.~U. Bajwa, J.~Haupt, A.~M. Sayeed, and R.~Nowak, ``Compressed channel
  sensing: A new approach to estimating sparse multipath channels,''
  \emph{Proc. of the IEEE}, vol.~98, no.~6, pp. 1058--1076, Jun. 2010.

\bibitem{SinPonMad:Multi-Gigabit-communication:-the-ADC-bottleneck1:09}
J.~{Singh}, S.~{Ponnuru}, and U.~{Madhow}, ``Multi-gigabit communication: the
  {ADC} bottleneck,'' in \emph{Proc. IEEE Int. Conf. on Ultra-Wideband}, Sep.
  2009, pp. 22--27.

\bibitem{delphi}
\BIBentryALTinterwordspacing
{Delphi Electronically Scanning Radar}. [Online]. Available:
  \url{https://autonomoustuff.com/product/aptiv-esr-2-5-24v/}
\BIBentrySTDinterwordspacing

\bibitem{Max:Quantizing-for-minimum-distortion:60}
J.~{Max}, ``Quantizing for minimum distortion,'' \emph{IRE Transactions on
  Information Theory}, vol.~6, no.~1, pp. 7--12, Mar. 1960.

\bibitem{Jay:Digital-Coding-of-Waveforms:84}
N.~S. Jayant and P.~Noll, \emph{Digital Coding of Waveforms}.\hskip 1em plus
  0.5em minus 0.4em\relax Prentice-Hall, 1984.

\bibitem{Ran:Generalized-approximate-message:11}
S.~{Rangan}, ``Generalized approximate message passing for estimation with
  random linear mixing,'' in \emph{Proc. IEEE Int. Symp. on Inf. Theory}, Jul.
  2011, pp. 2168--2172.

\bibitem{MezNos:Efficient-reconstruction-of-sparse:12}
A.~{Mezghani} and J.~A. {Nossek}, ``Efficient reconstruction of sparse vectors
  from quantized observations,'' in \emph{Proc. Int. ITG Workshop on Smart
  Antennas}, Mar. 2012, pp. 193--200.

\bibitem{VilSch:Expectation-Maximization-Gaussian-Mixture-Approximate:13}
J.~P. {Vila} and P.~{Schniter}, ``Expectation-maximization {Gaussian}-mixture
  approximate message passing,'' \emph{IEEE Trans. on Signal Process.},
  vol.~61, no.~19, pp. 4658--4672, Oct. 2013.

\end{thebibliography}

\end{document}